\documentclass[a4paper,11pt]{article}
\usepackage{slashed}
\usepackage{jheppub} 
\usepackage{lineno}
\usepackage{amsmath}
\usepackage{physics}
\usepackage{amsfonts}
\usepackage{amssymb}
\usepackage{mathtools}

\newcommand{\de}{\operatorname{d}\!}

\newcommand{\eqndot}{\, . }
\newcommand{\eqncom}{\, , }

\newcommand{\YM}{{\mathrm{\scriptscriptstyle YM}}}

\DeclareMathOperator{\cder}{D}
\newcommand{\scal}{\phi}
\newcommand{\scalc}{\scal^{\text{cl}}}
\newcommand{\scalq}{\tilde{\scal}}
\newcommand{\ferm}{\psi}
\newcommand{\aferm}{\bar{\ferm}}

\renewcommand{\digamma}{\Psi}

\newcommand{\sfrac}[2]{{\textstyle\frac{#1}{#2}}}
\newcommand{\half}{\sfrac{1}{2}}

\arxivnumber{2406.13741} 

\title{\boldmath Integrable Conformal Defects in $\mathcal{N}=4$ SYM}

\author{Marius de Leeuw}
\author{and Adolfo Holguin}
\affiliation{School of Mathematics,
Trinity College Dublin,
Dublin, Ireland}

\emailAdd{holguina@tcd.ie}
\emailAdd{deleeuwm@tcd.ie}
\abstract{In this paper we classify integrable conformal defects in $\mathcal{N}=4$ SYM theory for which the scalar fields pick up a non-trivial vacuum expectation value. Defects of this form correspond to Dirichlet boundary conditions that have a pole at the defect. These set-ups typically appear on the field theory side of probe brane set-ups in the AdS/CFT correspondence. We show that such defects, for any codimension, are related to fuzzy spheres. We discuss the properties of the different possible fuzzy spheres that can appear and present the corresponding Matrix Product States. We furthermore set-up the quantum field theoretic framework by computing the mass matrix and finding the propagators.}

\begin{document}
\maketitle
\flushbottom

\section{Introduction} \label{sec:intro}
The dynamics of defects in conformal field theories is an important subject describing a range of phenomena, from boundaries or interfaces to heavy extended probes such as monopoles and strings. The presence of a conformal defect generically breaks the conformal symmetry of the original system allowing for a much richer structure of correlation functions. Determining the conformal data of a defect CFT is therefore an interesting but mostly intractable problem. A perfect toy model for this problem arises from studying defects in planar $\mathcal{N}=4$ super Yang-Mills in four dimensions. The Dirichlet boundary conditions can be generalized to allow for solutions that exhibit poles at the defect \cite{Gaiotto:2008sa}. In that case the scalar and possibly the gauge fields pick up a non-trivial vacuum expectation value. At weak coupling one needs to expand the action around a different classical solution that is caused by the presence of the defect. Because of this, operators obtain non-trivial one-point functions already at classical level. Operators can also obtain non-trivial one-point functions in other defect set-ups, but these will only be non-zero at the quantum level  \cite{Kristjansen:2020mhn}. 
    The tree-level one-point functions of primary operators have an interpretation in terms of the integrable spin chain description as overlaps with a spin chain boundary state. A lot progress has been made in understanding a particular class of spin chain boundary states, the so-called \textit{integrable boundary states} \cite{Ghoshal:1993tm, Piroli:2017sei}. This class of states is characterized by certain selection rules that lead to remarkably simple overlaps with on-shell Bethe states. A full classification of such states even for the simplest integrable Hamiltonians is still lacking, but the states appear to fall into families of valance-bond states (VBS), matrix product states (MPS), and cross-cap states. This problem is believed to be intimately related to the classification of solutions to the boundary Yang-Baxter equation. Thus a defect in $\mathcal{N}=4$ SYM is understood to be integrable if its corresponding spin chain boundary state is an integrable boundary state. In general it is difficult to check whether this integrability is preserved beyond weak coupling, outside of supersymmetric cases where a partial classification is possible using the dual string description where the integrable boundary states are expected to be described by boundary conditions of the sigma model SCFT \cite{Mann:2006rh, Dekel:2011ja}.

The states that arise in defect set-ups of the planar $\mathcal{N}=4$ are usually of MPS type. Since such states are also interesting on their own right from the point of view of quantum integrable chains, classifying them is an important problem regardless of whether their integrability is broken by strong coupling effects. Having a weak-coupling description of the allowed boundary states is also useful for testing non-pertubative aspects of the $AdS/CFT$ correspondence, since one expects that the allowed boundary conditions for the string match on both sides of the duality and very little is known about integrable boundary conditions beyond those preserving half of the supersymmetries of the model. If one insists that the integrability structure does indeed persist to strong coupling, the weak coupling data can be used as a seed for bootstrapping finite coupling defect CFT data \cite{Jiang:2019xdz}.

In this paper, we consider the classification problem of integrable MPS arising from conformal defect set ups in $\mathcal{N}=4$ SYM, focusing on the $SO(6)$ sector of the model. By studying a simpler condition necessary for integrability of the MPS, we find that the allowed states are necessarily associated with non-commutative spheres. This allows for two new candidate integrable MPS, the fuzzy $S^3$ and fuzzy $S^5$. We then give evidence for their integrability, and provide a formula for the on-shell overlap for the fuzzy $S^3$ MPS in terms of Baxter-$Q$ functions in the $SU(2)$ sector. The fuzzy $S^5$ codimension-2 defect is identified with a corresponding half-BPS supersymmetric defect which is expected to be integrable from the classification of half-BPS integrable boundaries of the worldsheet sigma model. Our analysis thus captures all known half-supersymmetric integrable boundaries of the sigma model, and provides the possibility of additional nonsupersymmetric integrable D-brane configurations. 

Moreover, we show that the integrable MPS states can be used to describe defects with different codimensions. So far, most exact results have been derived for defects of codimension one. We will set-up the quantum field theory framework for integrable defect set-ups in higher codimensions as well. We present the propagators and compute the mass matrix. This will open up the possibility to obtain exact results and do explicit calculations for general codimension. Of particular interest is the codimension two case where a lot of progress has been made by the boundary conformal bootstrap program. 

The paper is structured as follows. In section 2 we review the details of the $\mathcal{N}=4$ SYM theory relevant to defect set-ups. In section 3 we discuss the relation between defects and integrable boundary states, and classify the matrix product states annihilated by the charge $Q_3$ of the integrable chain. In section 4 we outline the general construction of the fuzzy spheres of dimension up to five. In section 5 we reproduce various one point functions of all fuzzy sphere defects in the $SU(2)$ sector. We then discuss the spectrum of fluctuations around the defect solutions, focusing on the case of the fuzzy $S^3$, and give a general discussion of the propagators for all codimensions. In section 8 we give a discussion of the possible holographic interpretations of our results, matching the existing results for supersymmetric boundary conditions. We also provide appendices reviewing the details of the representation theory of $SO(n)$, the diagonalization of the mass matrix, and the dimensional reduction of spinors needed for the higher codimension defect set-ups. Finally we conclude with some comments on future directions and generalizations.

\section{Field theoretic set-up}

\paragraph{Field content and Action}
The field content of $\mathcal{N}=4$ SYM theory consists of a four-dimensional gauge field $A_\mu$, six real scalars $\phi_i$ and four four-dimensional Majorana fermions $\psi$. The action of $\mathcal{N}=4$ SYM theory is given by
\begin{align}
 \label{SYMaction}
  S_{{\mathcal N}=4}=&~\frac{2}{g_\YM^2}\int \de^4x\,\tr\biggl[ -\frac{1}{4}F_{\mu\nu}F^{\mu\nu}-\frac{1}{2}\cder_\mu\scal_i\cder^\mu\scal_i \nonumber\\
  &\qquad+\frac{i}{2}\aferm\Gamma^\mu\cder_\mu\ferm +\frac{1}{2}\aferm\Gamma^i\comm{\scal_i}{\ferm}+\frac{1}{4}\comm{\scal_i}{\scal_j}\comm{\scal_i}{\scal_j}\biggr]\eqncom
\end{align}
where the field strength $F_{\mu\nu}$ and the covariant derivatives $\cder_\mu$ are defined in the usual way 
\begin{equation}
 \begin{aligned}
  & \hspace{.5cm}F_{\mu\nu}=\partial_\mu A_\nu-\partial_\nu A_\mu-i\comm{A_\mu}{A_\nu}\eqncom\\
  \cder_\mu\scal_i&=\partial_\mu\scal_i-i\comm{A_\mu}{\scal_i}\eqncom \hspace{0.5cm}
  \cder_\mu\psi=\partial_\mu\psi-i\comm{A_\mu}{\psi}\eqndot
 \end{aligned}
\end{equation}

\paragraph{Equations of Motion}

We are interested in solutions to the equations of motion for the scalar fields and the gauge fields. They are given by
\begin{align}
&D^2 \phi_i = \sum_{j} [\phi_j,[\phi_j,\phi_i]],
&& D_\mu F^{\mu\nu} = \sum_i  [\phi_i , D^\nu \phi_i].
\end{align}
We will consider conformal defects. These are defects that preserve part of the conformal algebra including scaling transformations. Because of this, conformal defects are restricted to coordinate hyperplanes $x_i=0$ and are characterised by their (co-)dimension. 
For any codimension, we can go to spherical or cylindrical coordinates in the space perpendicular to the defect. Due to the symmetry of our set-up we can assume that the scalar fields only depend on the radial coordinate $\phi_i = \phi_i(z)$. Setting the gauge field to zero, our equations of motion reduce to
\begin{align}
\partial^2_z \phi_i + \frac{3-p-1}{z}\partial_z \phi_i = \sum_{j} [\phi_j,[\phi_i,\phi_j]],
\end{align}
where $3-p$ is the codimension of our defect. We find that the solution to this is
\begin{align}
\phi_i = \frac{t_i\oplus 0_{N-k}}{z},
\end{align}
where $t_i$ are $k\times k$ matrices that satisfy the equation
\begin{align}\label{t-matrix equation}
\sum_{j} [t_j,[t_i,t_j]]  = t_i \sqrt{p}.
\end{align}
Apart from the codimension-3 case, the factor of $\sqrt{p}$ can be absorbed in the normalization of $t_i$.  Now we consider the possibility of having non-zero gauge fields. Due to the symmetries of the solutions it is natural to fix a gauge for the classical gauge field profile such that $A_z=0$, which leaves the equations for the scalar fields unchanged. If we insist on having scalar field profiles that manifest the scaling symmetry then the right-hand side of the equation of motion for the field strength will vanish, so the gauge connection will be flat. The only interesting possibilities are then those which have non-trivial topology for the gauge field. For example, in codimension 3 there is a non-trivial 2-cycle in the directions tranverse to the defect on which we can support either electric or magnetic flux on. The gauge connection can be chosen to be diagonal. It should also be noted that the classical equation of motions for the scalars force the fields to commute with themselves, and charge quantization forces the scalars to be simultaneously diagonal with the gauge connection. This means that the only possible solutions are the background of electric, magnetic, and dyonic line defects which correspond to combinations of Wilson and 't Hooft lines. In codimension 2 we can have a non-trivial gauge field configuration along the defect itself. The only consistent possibility there is for the gauge field to be of the form $A_\psi= \alpha d\psi$ for a constant $\alpha$, where $\psi$ is the polar angle on the plane tranverse to the surface defect. This breaks rotational symmetry along the transverse direction to the defect, but the  In that case we should not assume that the scalar fields are independent of the angle $\psi$ and instead we should look for solutions that are covariantly constant $D_\psi \phi^i=0$ which once again forces the scalar field to commute with the scalars. These kind of solutions are described by Gukov-Witten defects. For codimension one there is no two-cycle for which we can thread flux without breaking the conformal symmetry along the defect. This means that the only non-trivial solutions with manifest conformal symmetry and non-commuting scalar matrices have vanishing gauge fields.

\paragraph{Gauge-Fixing}
To determine the masses of the various fields we need to expand the action to quadratic order around the defect solution 
\begin{equation}
\phi_i= \phi_i^{cl}+ \tilde{\phi}_i.
\end{equation}
Following the conventions in \cite{Alday:2009zm, Buhl-Mortensen:2016pxs,Buhl-Mortensen:2016jqo}, we will work in a gauge where
\begin{equation}
\partial^{\mu} A_\mu + i[\tilde{\phi}_i, \phi_i^{cl}]=0.
\end{equation}
This is implemented by adding the following ghost terms to the action
\begin{equation}
\begin{aligned}
S_{gh}= \frac{2}{g_{YM}^2} \int d^4x \tr\bigg[&\bar{c}\left( \partial_\mu D^\mu c -[\phi_i^{cl}, [\phi_i^{cl}+ \tilde{\phi}_i, c]]\right)- \frac{1}{2}(\partial_\mu A^\mu)^2 \\ &+ i [A^\mu, \tilde{\phi}_i]\partial_\mu \phi_i^{cl}+
i [A^\mu, \partial_\mu \tilde{\phi}_i] \phi_i^{cl}+ \frac{1}{2}[\phi_i^{cl},\tilde{\phi}_i]^2\bigg].
\end{aligned}
\end{equation}
\paragraph{Quadratic Action}
After the ghost terms to the action and simplifying, the action becomes 
\begin{equation}
S_{\mathcal{N}=4}+ S_{gh}= S_{\text{kin}}+ S_{m,b}+ S_{m,f}+ S_{\text{cubic}}+ S_{\text{quartic}}.
\end{equation}
The kinetic terms in this gauge are
\begin{equation}
S_{\text{kin}}= \frac{2}{g^2_{YM}} \int d^4 x \tr\bigg[\frac{1}{2} A_\nu \partial^{\nu}\partial^\mu A_\mu+ \frac{1}{2} \tilde{\phi}_i \partial_\mu \partial^\mu \tilde{\phi}_i+ \frac{i}{2}\bar{\psi} \gamma^\mu \partial_\mu \psi+ \bar{c}\, \partial_\mu \partial^\mu c \bigg],
\end{equation}
the bosonic mass terms are 
\begin{equation}
\begin{aligned}
S_{m,b}= \frac{2}{g^2_{YM}} \int d^4 x \tr&\bigg[  \frac{1}{2}[\phi^{cl}_i, \phi^{cl}_j][\tilde{\phi}_i, \tilde{\phi}_j]+ \frac{1}{2}[\phi^{cl}_i, \tilde{\phi}_j][\phi^{cl}_i, \tilde{\phi}_j]+ \frac{1}{2}[\phi^{cl}_i, \tilde{\phi}_j][\tilde{\phi}_i, \phi^{cl}_j]\\
&+\frac{1}{2}[\phi^{cl}_i, \tilde{\phi}_i][\phi^{cl}_j, \tilde{\phi}_j]+ \frac{1}{2}[A_\mu, \phi_i^{cl}][A^\mu, \phi_i^{cl}]+ 2i [A^\mu, \tilde{\phi}_i]\partial_\mu \phi_i^{cl} \bigg],
\end{aligned}
\end{equation}
and the fermion and ghost mass terms are
\begin{equation}
S_{m,f} = \frac{2}{g^2_{YM}} \int d^4 x \tr\bigg[\frac{1}{2} \bar{\psi}\, G^i[\phi_i^{cl},\psi ]- \bar{c}[\phi_i^{cl},[\phi_i^{cl},c]] \bigg].
\end{equation}
Here $G^i$ are Clebsch-Gordon coefficients coupling the scalar and fermion $SO(6)_R$ flavor indices.

\paragraph{Interaction terms}
For completeness and future reference, let us also list the interaction terms. The cubic interaction is given by
\begin{multline}
  S_{\text{cubic}}
    = \frac{2}{g_\YM^2}\int \de^4 x \tr\biggl[
        i[A^\mu,A^\nu]\partial_\mu A_\nu
        +[\scalc_i,\scalq_j][\scalq_i,\scalq_j]
        +i[A^\mu,\scalq_i]\partial_\mu\scalq_i
        +[A_\mu,\scalc_i][A^\mu,\scalq_i]\\
        +\frac{1}{2}\aferm\gamma^\mu[A_\mu,\ferm]
        +\sum_{i=1}^3\frac{1}{2}\aferm G^i[\scalq_i,\ferm]
        +\sum_{i=4}^6\frac{1}{2}\aferm G^i[\scalq_i,\gamma_5\ferm]
        +i(\partial_\mu\bar{c})[A^\mu,c]
        -\bar{c}[\scalc_i,[\scalq_i,c]]\biggr]
        \label{eq:cubic},
\end{multline}
and the quartic interaction term is 
\begin{align}
 S_{\text{quartic}}=\frac{2}{g_\YM^2}\int \de^4x\,\tr\biggl[ 
 \frac{1}{4}\comm{A_\mu}{A_\nu}\comm{A^\mu}{A^\nu}
 + \frac{1}{2}\comm{A_\mu}{\tilde{\scal}_i}\comm{A^\mu}{\tilde{\scal}_i}
 +  \frac{1}{4}\comm{\tilde{\scal}_i}{\tilde{\scal}_j}\comm{\tilde{\scal}_i}{\tilde{\scal}_j}
 \biggr].
\end{align}

\section{Integrable defects}

\paragraph{Defects and vevs}
Defects in conformal field theory are modifications of a theory restricted to a lower dimensional subspace, with conformal defects being those modifications which preserve conformal symmetry along the defect. There are many ways of describing conformal defects, for instance by coupling the original system to a conformal theory supported on a codimension $p$ hyperplane or by the insertion of a non-local operator. One useful perspective for conformal defects in weakly coupled systems is to describe the effect of the defect insertion as a boundary condition for the bulk fields along the defect. If the system is weakly coupled the leading order description of the defect is captured by solving the classical equations of motion of the model with a singularity along the defect. The profile of the fields near the fields is fixed by conformal symmetry to be of the form
\begin{equation}
\phi^{cl}(z)\sim \frac{C_{\phi}}{z^{\Delta_{\phi}}}+ \dots,
\end{equation}
where $z$ is the coordinate normal to the defect, and $\Delta_{\phi}$ is the scaling dimension of the field $\phi$. The classical profiles of the fields lead to non-zero vacuum expectation values of operators which modify correlation functions of local operators. For example the tree-level one-point function of local operators in a defect conformal field theory is captured by the classical profile of the field
\begin{equation}
\langle \mathcal{O}(x)\rangle_{\text{tree-level}}= \mathcal{O}^{\,cl}(x)=\frac{C_{\mathcal{O}}}{z^{\Delta_{\mathcal{O}}}}.
\end{equation}
Quantum corrections are taken into account by quantizing the model around the classical solution sourced by the defect. In many cases there are additional corrections arising from interactions between bulk fields and the defect degrees of freedom \cite{Gaiotto:2008sa}.

The relevant defects in $\mathcal{N}=4$ SYM are solutions to the equations of motion for the fields with a prescribed singularity along the defect. For example solutions with non-diagonal scalar field profiles arise when the gauge fields and fermions are set to zero, the simplest family being of the form

\begin{equation}
\phi_i(z)= \frac{t_i\oplus 0_{N-k}}{z},
\end{equation}
with $t_i$ solving \ref{t-matrix equation}. The tree-level one-point functions of a generic scalar operator are obtained by substituting the classical profile into the operator

\begin{equation}
\Psi^{i_1 i_2 \dots i_L}\langle\tr_N\left[\phi_{i_1}(z)\phi_{i_2}(z)\dots \phi_{i_L}(z) \right]\rangle_{\text{tree-level}}= \frac{1}{z^L}\Psi^{i_1 i_2 \dots i_L}\tr_k\left[t_{i_1}t_{i_2}\dots t_{i_L} \right].
\end{equation}

\paragraph{Matrix Product States}
Defects of the type discussed above give rise to spin chain matrix products when computing one-point functions. More precisely an operator in planar $\mathcal{N}=4$ SYM can be associated with a wavefunction for the $PSU(2,2|4)$ XXX chain at weak coupling. The one point function of such an operator can be interpreted as in terms on an overlap with a matrix product state:
\begin{equation}
\langle\mathcal{O}_{\Psi}(z) \rangle= \frac{1}{z^L}\Psi^{i_1 i_2 \dots i_L}\tr_k\left[t_{i_1}t_{i_2}\dots t_{i_L} \right]=\frac{\langle \Psi | \text{MPS} \rangle}{z^L}.
\end{equation}  
At strong coupling this overlap has the interpretation of an overlap between a closed string state and a boundary state associated to defect brane in the bulk $AdS$ space. This analogy persists to weak coupling and the MPS can be thought of as a channel rotation of a boundary condition for the spin chain, so we will continue refer to such MPS as boundary states. Generic choices of boundary conditions will break the integrability of the spin chain, and only a very special class of boundary states preserve integrability. Such integrable states satisfy various selection rules as a consequence of the boundary Yang-Baxter equation, and lead to simpler overlaps. For this reason it is interesting to classify integrable boundary states, for instance in the $SO(n)$ $XXX$ spin chain. In particular, in $\mathcal{N}=4$ SYM, the $SO(6)$ sector is closed at one-loop level and as such, any MPS corresponding to an integrable defect has to give rise to an integrable state on the $SO(6)$ spin chain.

\paragraph{Integrability condition}
A boundary state, and by extension the defect that is associated with it, is called integrable if it annihilated by an infinite number of conserved charges \cite{Ghoshal:1993tm, Piroli:2017sei}
\begin{equation}
Q_s | \mathcal{B} \rangle=0.
\end{equation}
For the type of spin chains that we will consider the set of conserved charges will be the set of parity odd charged $Q_{2k+1}$.  This is encoded in the following condition

\begin{align}
t(u) |\mathcal{B}\rangle = \Pi t(u) \Pi |\mathcal{B}\rangle ,
\end{align}
with $t(u)$ being the transfer matrix of the model, and $\Pi$ the space parity operator 
\begin{equation}
\Pi | i_1 i_2 \dots i_L\rangle= |i_L i_{L-1}
\dots i_1\rangle 
\end{equation}
Checking this condition for an arbitrary boundary state is difficult in general. For the class of defects that we are studying the boundary state will be of matrix product state type, where the condition has been checked in some cases by performing certain similarities transformations on the transfer matrices. Here we will take a different approach, and instead check the action of a single conserved charge on a general matrix product state, and classify the kinds of matrix product states that are annihilated by it. This is not sufficient to guarantee integrability of the defect a priori, but the structure of the equation will turn out to be the same as the equations of motion of the fields of $\mathcal{N}=4$ SYM. It will turn out that there only are two new families of the solutions to the equations $Q_3 | \text{MPS}\rangle=0$ that do not fall into the known families of integrable MPS; the fuzzy $S^3$ and fuzzy $S^5$ MPS. We will later give evidence of their integrability and compute overlaps with some Bethe states.
\paragraph{Solving the integrability condition}
A necessary condition for a spin chain state $|\mathcal{B}\rangle $ to be integrable is that $Q_3$ annihilates it
\begin{align}
    Q_3|\mathcal{B}\rangle =0.
\end{align}
Let us now work out what this condition implies for a Matrix Product State on the $SO(6)$ spin chain. For the $SO(N)$ spin chain, $Q_3$ takes the form 
\begin{align}
Q_3 = \sum_{n=1}^L   (Q_3)_{i' j' k'}^{ijk} (E^{i'}_i)_{n-1}(E^{j'}_j)_{n}(E^{k'}_k)_{n+1},
\end{align}
where $(E^i_j)_n$ is the matrix unity acting on the $n$th site and we sum over repeated indices. The coefficients of $Q_3$ are given by
\begin{align}
 (Q_3)_{i' j' k'}^{ijk}= &~ \delta^{ij} \delta^k_{i'} \delta_{j'k'}-\delta^i_{k'}\delta^{jk}\delta_{i' j'}\\
 &+ \Delta\left(\delta^{jk} \delta^i_{j'}\delta_{i' k'}-\delta^{ik}\delta^j_{i'}\delta_{j'k'} +\delta^{ik}\delta^j_{k'}\delta_{i'j'}-\delta^{ij}\delta^k_{j'}\delta_{i'k'} \right)\\
&+\Delta^2\left( \delta^i_{k'}\delta^j_{i'}\delta^k_{j'}-\delta^i_{j'}\delta^j_{k'}\delta^k_{i'}\right)
\end{align}
with $\Delta = \frac{2}{N-2}$. For $SO(6)$ we have $\Delta=2$. 
Consider now a collection of matrices $t_i$ which we use to construct a MPS of the form
\begin{align}
|\mathcal{B} \rangle = \sum_{\{i_k\}}\tr\left[t_{i_1}\dots t_{i_{L}}\right]|i_1, i_2, \dots, i_L\rangle.
\end{align}
We will assume that the indices of $t_i$ transform under $SO(6)$ transformations as the scalar fields do.  This assumption turns out to be essential for the integrability of the MPS. Applying $Q_3$ on a set of three neighboring sites gives the following (see also  Appendix A of \cite{deLeeuw:2015hxa})
\begin{align}\label{Q3action1}
    \left(Q_3\right)^{lnm}_{ijk}\,t_{l} t_{n} t_{m} =&~ \delta_{ij}[t_k, t_s]t_s + \delta_{jk} t_s [t_i, t_s]- \delta_{ij} t_s t_k t_s+ \delta_{jk} t_s t_i t_s \nonumber\\
&+ 2 \delta_{ik} [t_s t_s, t_j] + 4 [t_j, t_k t_i].
\end{align}
Note that there are three different types of tensor structures in \eqref{Q3action1}: the first four terms only involve tracing over neighboring sets of spins, the fifth term involves tracing over next-to-nearest neighbors, and the last term involves permutations of all three sites. To simplify this expression further we can combine  all the terms with similar tensor structures. Doing this yields a slightly nicer expression for the action of $Q_3$ on three sites of the MPS:
\begin{align}\label{Q3action2}
    \left(Q_3\right)^{lnm}_{ijk}\,t_{l} t_{n} t_{m} =&~ \delta_{ij}\left([t_s,[t_s, t_k]] -t_s t_s t_k \right)- \delta_{jk}\left( [t_s, [t_s, t_i]]-t_i t_s t_s\right) \nonumber\\
&+ 2 \delta_{ik} [t_s t_s, t_j] + 4 [t_j, t_k t_i].
\end{align}
We see that the third term in \eqref{Q3action2} can generically not be canceled the other terms because they come from different tensors structures. Hence we must insist that it vanishes by itself
\begin{align}\label{eq:con1}
&    [t_s t_s, t_j]=0.
\end{align}
In other words, this implies that $t_s t_s$ commutes with all of the $t_j$. There are two possible solutions to this.
First, we can assume that all of the $t_i$ commute. This trivially satisfies condition \eqref{eq:con1} and $Q_3$ will annihilate the state after summing over all the sites of the chain. In this case we can choose all our matrices $t_i$ to be diagonal. The MPS greatly simplifies in this case and is known to be integrable \cite{DeLeeuw:2019ohp, Kristjansen:2023ysz, Ivanovskiy:2024vel}. 
Second, we consider the case where the $t_i$ do not commute. If we then consider the polynomial algebra $\mathbb{C}[\{t_i\}]$ generated by the $t_i$, we can say that $t_s t_s$ is a central element. Since $SO(6)$ acts on the $t_i$ as a vector $R_i^j t_j=t_i$ and $t_s t_s$ is invariant under this transformation (thinking of the $t_i$'s as vectors with the usual inner product), Schur's lemma implies that $t_s t_s$ is a multiple of the identity, or more generally a direct sum of multiples of the identity acting on the irreducible representation that make up $\mathbb{C}[\{t_i\}]$ as a vector space. 
Without loss of generality we can reabsorb phases in such a way that $ t_s t_s$ is real and work with the case where we have only one block. This condition implies the equation for a fuzzy sphere \cite{Ramgoolam:2001zx}:
\begin{align}\label{eq:fuzzysphere}
    \sum_{s=1}^m\,t_s t_s= R^2 \,\mathbb{I}.
\end{align}
The dimension of the fuzzy sphere is set by the number of non-zero matrices $t_i$, which we call $m$.  Note that if we dropped the assumption that $SO(6)$ acts on the MPS matrices we would obtain solutions to \eqref{eq:con1} that do not lead to integrable MPS such as the $S^2\times S^2$ solution in\cite{deLeeuw:2019sew}. What happens in that case is that the last term in \eqref{Q3action2} does not cancel after summing over sites. Rewriting the last term using commutator identities gives
\begin{align}
[t_j, t_k t_i]= t_i [t_j, t_k]- [t_i, t_j] t_k -[t_i, [t_j, t_k]]+ [[t_i,t_j], t_k],
\end{align}
and clearly the first two contributions vanish after imposing the peridicity of the trace.  To eliminate the remaining terms we need to use symmetry to constraint the possible tensor structures in the commutator $[t_i,t_j]$.  Since \eqref{eq:fuzzysphere} has a natural $SO(m)$ invariance we see that $J_{ij}=[t_i,t_j]$ has the right tensor structure to be a  generator of $SO(m)\subset SO(6)$. The $SO(6)$ symmetry assumption implies that the only consistent tensor structure in the problematic terms is of the form
\begin{align}
  [J_{ij}, t_k]\propto \delta_{jk} t_i - \delta_{ik} t_j,
\end{align}
which guarantees that $Q_3$ annihilates the matrix product state. This relation is not possible for the $SU(2)\times SU(2)$ symmetric MPS since there are $SO(m)\subset SO(6)$ transformations that do not leave the MPS invariant. 

\section{Fuzzy Spheres}

We find that all the integrable defect solutions are given by fuzzy spheres. We give the explicit construction of the fuzzy spheres that appear here. 
There are four cases of fuzzy spheres that are relevant for the $SO(6)$ chain; the fuzzy $S^2$ case was studied in the context of the D3-D5 dCFT \cite{Constable:1999ac,deLeeuw:2015hxa} and the fuzzy $S^4$ appeared in the D3-D7 setup \cite{Castelino:1997rv,Constable:2001ag, deLeeuw:2016ofj}. The remaining cases are the odd dimensional fuzzy spheres $S^3$ and $S^5$. Odd-dimensional fuzzy spheres are more complicated, but their construction can be understood as a kind of embedding into an even dimensional fuzzy sphere in one higher dimension. The fuzzy $S^3$ is the simplest case, and in practice the other odd-dimensional spheres can be found in the same way; the main technical challenge is to construct projectors into a particular set of reducible representations of $SO(2k)$ that appear when we decompose the spinor representation of $SO(2k+1)$. 

\paragraph{Fuzzy $S^2$'s}
The fuzzy 2-sphere is the simplest kind to construct. The coordinates $J_i$ are $k\times k$ matrices that satisfy the commutation relations for $SU(2)$:
\begin{equation}
[J_i, J_j]= i \epsilon_{ijk} J_k.
\end{equation}
The square of the radius of the sphere is related to the quadratic Casimir $\sum J_i^2= j(j+1)$ where $k= 2j+1$. 

\paragraph{Fuzzy $S^3$ and $S^4$}
Fuzzy spheres of odd dimensions are more subtle to construct as opposed to their even dimensional analogs. For the fuzzy $S^3$ we should begin with a fuzzy $S^4$ and proceed to foliate it by $S^3$ slices of varying radius. This amounts to finding a decomposition of the defining representation for the fuzzy $S^4$ into direct sums of $SO(4)$ reducible representations. To construct a fuzzy $S^4$ we start with the spinor representation of $SO(5)$ and its associated gamma matrices $\Gamma_\mu$ with $\mu=1,2,\dots, 5$. From these we can construct $G_\mu$ matrices that satisfy the equation
\begin{equation}
	\sum_\mu G_\mu G_\mu= R^2 \,\mathbb{I}_d,
\end{equation}
where $d$ is the dimension of the matrices $G_
\mu$. Their construction is given for instance in \cite{ Ramgoolam:2001zx, deLeeuw:2016ofj}, 

\begin{equation}
	G_\mu = \left(\Gamma_\mu \otimes 1\otimes 1\dots \otimes 1+ 1\otimes \Gamma_\mu \dots \otimes + \dots\right)_{\text{sym}},
\end{equation}
where we project the tensors onto their fully symmetric components. The commutators of the matrices $G_{\mu}$ will then satisfy the algebra of $SO(5)$ and will act on an irreducible representation $Sym\left(V_5^{\otimes n} \right)$. We can ask how this space decomposes as representations of  $SO(4)\simeq SU(2)_L\times SU(2)_R\subset SO(5)$. This symmetry breaking pattern slices the $S^4$ into $S^3$ slices whose radii depend on the dimension of the corresponding $SO(4)$ representations. 
\begin{equation}
	Sym(V_5^{\otimes n})= \bigoplus_{l=0}^n V_{\frac{n-l}{2}}\otimes V_{\frac{l}{2}},
\end{equation}
where $V_{\frac{n-k}{2}}$ is the spin $j=\frac{n-k}{2}$ representation of $SU(2)$. Each pair of factors corresponds to a $S^3$ slice of the fuzzy $S^4$ whose radius is given by their eigenvalue under $R^2-G_5 G_5$. The prescription of \cite{Ramgoolam:2001zx} is to a pair of subspaces with maximal radius; this makes sense for the following reason. For each $l$ we can associate a fuzzy $S^3$ via the spaces 
\begin{equation}
	\left(V_{\frac{n-l}{2}}\otimes V_{\frac{l}{2}}\right)\oplus \left(V_{\frac{l}{2}}\otimes V_{\frac{n-l}{2}}\right).
\end{equation}
We can decompose the tensor products further 
\begin{equation}
	\frac{n-l}{2}\otimes \frac{l}{2}= \left(\frac{n}{2}-l\right)\oplus  \left(\frac{n}{2}-l-1\right)\dots\oplus \frac{n}{2};
\end{equation}
essentially what we are doing here is decomposing $SO(4)$ representations in terms of the diagonal $SO(3)\subset SO(4)$ which is the local Lorentz group on $S^3$. Each of the factors in the tensor product correspond to sections of a bundle of this fuzzy sphere, and we can see that the lowest spin mode we see is $\frac{n}{2}-l$. In order to see see all the modes we should take the $2l= n\pm 1$, that way the tensor product decomposition starts at spin one-half and ends with spin $n/2$. Now the task is to construct a projector $P_\mathcal{R}$ into the representation 
\begin{equation}
	\mathcal{R}= \left(V_{\frac{n+1}{4}}\otimes V_{\frac{n-1}{4}}\right)\oplus \left(V_{\frac{n-1}{4}}\otimes V_{\frac{n+1}{4}}\right).
\end{equation}
This is equivalent to finding the eigenvectors of $G_5$ with eigenvalues $\pm 1$ and projecting into that eigenspace. In practice this may be done by taking the results in \cite{deLeeuw:2016ofj} and restricting the sums to the two terms with $2j= n\pm 1$, for a nice choice of vacuum state. More generally the projector is given by 
\begin{equation}
P_{\mathcal{R}}\sim	\prod_{i}\left( G_5 G_5- R_i\right),
\end{equation}
with the product taken over the eigenvalues of $G_5$ that are not equal to one. 

To find the mass terms for quadratic fluctuations we will need to decompose the color indices into irreducible representations of $SO(4)$. For an classical solution with a single block of size $d_G$, we can split the color indices for a flunctuation into blocks

\begin{equation}
\delta \phi= \begin{pmatrix} \delta \phi_{m,m'}&& \delta \phi_{m, a'}\\ 
\delta \phi_{ a, m'}&& \delta \phi_{a, a'}\end{pmatrix}.
\end{equation}
The only modes that get mass terms are those in the first block and the off-diagonal blocks. The off-diagonal fluctuations transform in $2(N-d_G)$ copies of the $\mathcal{R}_+ \oplus\mathcal{R}_- $ representation while the decomposition of the adjoint block is more complicated. The $ \delta \phi_{m,m'}$ block needs to be decomposed further into four square blocks which we can associate with the spaces $\text{End}(\mathcal{R}_+)$, $\text{Hom}(\mathcal{R}_+, \mathcal{R}_-)$, $\text{Hom}(\mathcal{R}_-, \mathcal{R}_+)$, and $\text{End}(\mathcal{R}_-)$. In terms of $\mathfrak{su}(2)\times \mathfrak{su}(2)$ spins we have that $\text{End}(\mathcal{R}_+)\cong \mathcal{R}_+\otimes \mathcal{R}_+\cong (\frac{n+1}{4}\otimes\frac{n+1}{4}, \frac{n-1}{4}\otimes\frac{n-1}{4} )$; similarly for $\text{Hom}(R_+,R_-)\cong R_+\otimes R_-$.  So we have the decomposition
\begin{equation}\label{dimensions}
\begin{aligned}
\text{End}(\mathcal{R}_+)&\cong \bigoplus_{j_1=0}^{\frac{n+1}{2}}  \bigoplus_{j_2=0}^{\frac{n-1}{2}}(j_1,j_2)\\
\text{Hom}(R_+,R_-)&\cong \bigoplus_{j_1=\frac{1}{2}}^{\frac{n}{2}}  \bigoplus_{j_2=\frac{1}{2}}^{\frac{n}{2}}(j_1,j_2).
\end{aligned}
\end{equation}
The decomposition for the remaining blocks is obtained by exchanging the role of $j_1$ and $j_2$. Note that we used the fact that $SU(2)$ representations are self-dual to identify $\mathcal{R}_+$ and  $\bar{\mathcal{R}}_+$.
\subparagraph{Fuzzy $S^5$}
The construction for a general fuzzy $S^5$ is completely parallel to the fuzzy $S^3$ case. Given the seven-dimensional Gamma matrices $\Gamma_a$ one can build $G_a$ matrices by taking an $n$-fold symmetrized tensor product.
\begin{equation}
	G_a= \left(\Gamma_a \otimes 1\otimes 1\dots \otimes 1+ 1\otimes \Gamma_a \dots \otimes + \dots\right)_{\text{sym}}.
\end{equation}
 Their commutators are generators of $SO(7)$ in the $\text{Sym}(V_8^{\otimes n})$ irreducible representation. The matrices $G_a$ with $a=1,\dots, 7$ form a fuzzy $S^6$.  Together with their commutators the matrices $L_{a8}=G_a$ satisfy the algebra of $SO(8)$. To obtain a fuzzy $S^6$ we decompose the seven dimensional spinor representation into six dimensional spinors $V_8= V_4\oplus \bar{V}_4$ and use this to reduce $\text{Sym}(V_8^{\otimes n})$ into irreducible representations of $SO(6)$. Within these we look for the subspace for which $\sum_{a=1}^6 G_a^2$ has its largest eigenvalue. We then decompose this subspace in terms of the sign of the eigenvalue of  $G_7$. We call these $\mathcal{R}_+$ and $\mathcal{R}_-$ and their weights are $(\frac{n}{2}, \frac{n}{2}, \frac{1}{2})$ and $(\frac{n}{2}, \frac{n}{2}, -\frac{1}{2})$. The restriction of $G_a$into the space $\mathcal{R}=\mathcal{R}_+\oplus \mathcal{R}_-$ which we call $\hat{G}_a$ will form a fuzzy $S^5$.

As with the fuzzy $S^3$, the fuzzy harmonics on the $S^5$ are obtained by decomposing the spaces $\text{End}(\mathcal{R}_{\pm})$ and $\text{Hom}(\mathcal{R}_{\pm}, \mathcal{R}_{\mp})$ into irreducible representations of $SO(6)$; this decomposition was carried out in \cite{Ramgoolam:2001zx}.
\section{One-point functions}
We will now collect some simple overlaps associated with the various fuzzy sphere matrix product states, focusing on the $SU(2)$ and Konishi operators. For the fuzzy $S^3$ MPS we are able to provide a formula for the overlap with arbitrary Bethe eigenstates in terms of Q-functions in the $SU(2)$ sector. Similar formulas exist for the full $SO(6)$ sector for the $S^3$ and $S^5$ MPS of arbitrary bond dimension but we postpone that analysis for future work. 
\subsection{Fuzzy $S^5$}
Because the defect has $SO(6)$ symmetry, its overlap with any BMN vacuum vanishes. Another way of seeing this is that the overlap is always a trace of an $SO(8)$ raising operator
\begin{equation}
\langle \text{MPS}| Z^L \rangle= \tr \;(\hat{G}_5+ i \hat{G}_6)^L\sim \tr\; (L_{58}+ i L_{68})^L.
\end{equation}
This means that only half-filling states above the BMN vacuum have non-zero overlaps. The overlap with the Konishi operator is proportional to the square of the radius of the fuzzy $S^5$
\begin{equation}
\langle \text{MPS}| \mathcal{K}\rangle= \tr\; \hat{G}_a \hat{G}_a= R_5(n)^2\times 2\times \left(\frac{1}{192}(n+1)(n+3)^3(n+5) \right).
\end{equation}
    The function $R_5(n)^2$ is related to a difference of $SO(7)$ and $SO(6)$ quadratic Casimirs \cite{Medina:2002pc}. This is because the matrices $G_a\propto L_{a8}$ are proportional to $SO(7)\subset SO(8)$ generators on each of the different $SO(6)$ irreducible blocks within $\text{Sym}(V_8^{\otimes n})$, which lets us rewrite the radius of the fuzzy sphere as:
 \begin{equation}
\sum_{a=1}^6 \hat{G}_a^2= 2 \left(\,C^{SO(7)}_2(\text{Sym}(V_8^{\otimes n}))-\frac{1}{2}C_2^{SO(6)}(\mathcal{R})= \,C^{SO(7)}_2(\text{Sym}(V_8^{\otimes n}))-C_2^{SO(6)}(\mathcal{R}_{\pm})\right).
\end{equation}
This gives that 
\begin{equation}
R_5(n)^2= n(n+6)-1,
\end{equation}
which is consistent with the fact that the fuzzy $S^5$ is the "equator" of a fuzzy $S^6$ of radius $n(n+6)$;
\begin{equation}
\sum_{a=1}^6 G_a^2= R_6(n)^2= 2\left(C_2^{SO(8)}\left(\frac{n}{2}, \frac{n}{2}, \frac{n}{2}, \frac{n}{2}\right)-C_2^{SO(7)}\left(\frac{n}{2}, \frac{n}{2}, \frac{n}{2}\right)\right)= n(n+6).
\end{equation}
\subsection{ $S^4$}

Now we collect some of the results for overlaps of Bethe states with the fuzzy $S^4$ MPS. 
For the BMN vacuum $\tr Z^L$ the overlap is 
\begin{equation}
	\langle \text{MPS}| 0\rangle = \tr G_5^L.
\end{equation}
Before giving value of this correlator we should give a few comments that will be relevant for later calculations. From our discussion about the fuzzy $S^3$ we know that $G_5$ for generic rank can be block diagonalized into $n$ blocks of size $j(n-j+2)$. By group theory arguments we can infer that $G_5$ can be taken to be a multiple of the identity on each block, and we see that the eigenvalues of are precisely the difference of the left and right $SU(2)$ spins under the $SO(4)$ decomposition
\begin{equation}
\begin{aligned}
	\tr G_5^L&= \sum_{l=0}^n (l+1)(n-l+1)(2l-n)^L\\
	&=\left[ 1^L+(-1)^L\right]\left[\frac{2}{L+3} B_{L+3}(-n/2)-\frac{(n+2)^2}{2(L+1)} B_{L+1}(-n/2) \right],
\end{aligned}
\end{equation}
where $B_L(x)$ are Bernoulli polynomials. The next simplest state is the Konishi state $|\mathcal{K}\rangle = \sum_i \tr \phi_i \phi_i$. 
\begin{equation}
\langle \text{MPS}| \mathcal{K}\rangle = \frac{1}{48}n(n+1)(n+2)(n+3)(n+4).
\end{equation}
For this case the radius may be related to quadratic Casimirs as
\begin{equation}
\sum_{I=1}^5 G_I G_I= 2\left( C^{SO(6)}_2\left(\frac{n}{2},\frac{n}{2},\frac{n}{2}\right)-C_2^{SO(5)}\left(\frac{n}{2},\frac{n}{2} \right)\right)
\end{equation}
By symmetry the overlaps with Bethe states in the $SU(2)$ sector will vanish. The overlap with general $SO(6)$ Bethe states was found in \cite{DeLeeuw:2019ohp}.

\subsection{ $S^3$}
For the fuzzy $S^3$ it is more convenient to introduce a fifth matrix $G_5$ and embed into a fuzzy $S^4$. To compute the correlators we simply need to restrict the traces to the subspaces with $2l-n=\pm 1$. Then we proceed just as with the fuzzy $S^4$ case except that no scalar field can be replaced by $G_5$. For example we can consider a choice of BMN vacuum $|\tilde{\Omega}\rangle=tr\left[\left(\phi_4+i\phi_5\right)^L\right]$, which has the following overlap with the fuzzy $S^3$ MPS

\begin{equation}
	\langle \text{MPS}|\tilde{\Omega} \rangle= \tr \hat{G}_4^L,
\end{equation}
where $\hat{G}_i$ is the restriction of the fuzzy $S^4$ matrices to the appropriate reducible representation of $SO(4)$. The matrix $\hat{G}_4$ can be then diagonalized and shown to be of the following form
\begin{equation}
\hat{G}_4= \text{diag}\left\{\overbrace{l,\dots l}^{2l\; \text{times}}, \overbrace{(l-1),\dots(l-1)}^{2l-2\; \text{times}}, \dots , \overbrace{-l,\dots -l}^{2l\; \text{times}} \right\},
\end{equation}
where $n= 2l-1$. The overlap with the BMN vacuum evaluates to 
\begin{equation}
\langle \text{MPS}|\tilde{\Omega} \rangle=2\left[ 1^L+(-1)^L\right]\times \sum_{k=1}^{\frac{(n+1)}{2}} k^{L+1}
\end{equation}
 In order to get a non-trivial one point function the value of $n$ in $\text{Sym}^n(V)$ must be odd. 
The overlap with the Konishi operator we get 
\begin{equation}
\langle \text{MPS}| \mathcal{K}\rangle= \tr \hat{G}_i \hat{G}_i= \left(n(n+4)-1 \right)\times \frac{(n+1)(n+3)}{2}.
\end{equation}
This can be obtained by using the Casimir relation 
\begin{equation}
\sum_{i=1}^4 \hat{G}_i \hat{G}_i= 2\left(C^{SO(5)}_2\left(\frac{n}{2}, \frac{n}{2} \right)-\frac{1}{2}C_2^{SO(4)}(\mathcal{R})\right)
\end{equation}
Unlike the fuzzy $S^4$ MPS, the fuzzy $S^3$ has non-vanishing overlaps in the $SU(2)$ sector, for example we can use the following complex scalar fields
\begin{equation}
\begin{aligned}
\tilde{Z}&= \phi_4+i \phi_5\\
\tilde{Y}&= \phi_1+ i \phi_6.
\end{aligned}
\end{equation}
The two-magnon overlap can be shown to be of the form
\begin{equation}
\begin{aligned}
\frac{C_n^{S^3} \left(u,-u\right)}{C_1^{S^3} \left(u,-u\right)}&=  \frac{1}{2}\sum_{j=\frac{1}{2}}^{\frac{n}{2}} \left(j+\frac{1}{2}\right)^{L}\sum_{m=-j}^{j} \mathcal{F}_{jm}(u)\\
\mathcal{F}_{jm}(u)&= \frac{\left(u^2+ \left(\frac{n+2}{2}\right)^2\right)\left(u^2+m^2\right)}{\left(u^2+\left(j+1\right)^2\right)\left(u^2+j^2\right)}
\end{aligned}.
\end{equation}
More generally the overlap for on-shell $SU(2)$ Bethe states takes the form\footnote{We thank Tamás Gombor for clarifying the generalization of the two-magnon formula.}

\begin{equation}
\frac{C_n^{S^3} \left(\{u_i\}\right)}{C_1^{S^3} \left(\{u_i\}\right)}=  \frac{1}{2}\sum_{j=\frac{1}{2}}^{\frac{n}{2}} \left(j+\frac{1}{2}\right)^{L}\sum_{m=-j}^{j} \prod_{i=1}^{\frac{M}{2}} \mathcal{F}_{jm}(u_i).
\end{equation}

\subsection{$S^2$}
The fuzzy $S^2$ is the simplest since the matrix elements of the matrices can be written down explicitly. For arbitrary $k\times k$ generators of $SO(3)$ the one point functions in the $SU(2)$ sector are given by \cite{Buhl-Mortensen:2015gfd}

\begin{align}
C_k^{S^2} \left(\left\{u_j\right\}\right) &= 
2^{L-1} C_2^{S^2}\left(\left\{u_j\right\}\right) 
\sum_{j=\frac{1-k}{2}}^{\frac{k-1}{2}}  j^L \prod_{i=1}^{\frac{M}{2}} 
\,\frac{u_i^2\left(u_i^2 + \frac{k^2}{4}\right)}{\left[u_i^2+(j-\half)^2\right]
\left[u_i^2+(j+\half)^2\right]}\, \\
 C_2 \left(\left\{u_j\right\}\right) &=2\left[
 \left(\frac{2\pi ^2}{\lambda }\right)^L\frac{1}{L}
 \prod_{j}^{}\frac{u_j^2+\frac{1}{4}}{u_j^2}\,\,\frac{\det G^+}{\det G^-}\right]^{\frac{1}{2}}.
\end{align}
There also exist exact functions for the overlap in the SU(3) and SO(6) sectors that are closed at one loop \cite{deLeeuw:2016umh,DeLeeuw:2018cal}.

\section{Mass Matrix}

To determine the masses of the various fields we need to diagonalize the quadratic mass terms in the action. 
This was done in detail for the fuzzy $S^2$ in \cite{Buhl-Mortensen:2016jqo} and for the fuzzy $S^4$ in \cite{Gimenez-Grau:2019fld}, and the analysis for a general fuzzy sphere is very similar except for technical details related to the representation theory of even dimensional orthogonal groups. We will explain the diagonalization procedure for the fuzzy $S^3$ in detail.  The most complicated part of the diagonalization of the mass matrix is the mixing problem of the longitudinal perturbations on top of the defect. In general the flavor indices of the modes decompose into transverse and longitudinal modes along the fuzzy sphere. The transverse fluctuations are simpler and do not mix in the flavor index space. The diagonalization of the longitudinal modes involves computing certain matrix elements of spin generators of higher dimensional orthogonal groups. 

Following the notation in \cite{Buhl-Mortensen:2016jqo, Gimenez-Grau:2019fld}, the commutators of the $G_I$  matrices for a $2k$ dimensional fuzzy spheres are related to $SO(2k+1)$ generators via  $L_{IJ}= -i[G_I, G_J]$, and the remaining generators of $SO(2k+2)$ are given by $L_{I, 2k+2}= G_I$.  Note that our definition of the $G$ matrices differs from theirs by a factor of $\frac{1}{2}$.  A $2k-1$ dimensional fuzzy sphere is obtained by restricting the $G_I$ to the appropriate reducible representations of $SO(2k)\subset SO(2k+1)\subset SO(2k+2)$. This amounts to restricting the range of $I$ from 1 to $2k-1$ and projecting into the subspace with largest eigenvalue for $\sum_{I=1}^{2k-1} G_{I}^2$. As a result the commutators  $L_{IJ}$ still satisfy the algebra of $SO(2k)$, but the restricted matrices $\hat{G}_I$ do not quite complete the algebra to $SO(2k+1)$ but should still be thought of as $SO(2k+2)$ generators.  We will also often encounter the adjoint action of $L_{I,J}$ on the fluctuations $\delta \phi_{n,m}$
\begin{equation}
\mathcal{L}_{IJ} \circ \delta \phi_{mn}= [L_{IJ}, \delta \phi_{mn}],
\end{equation}
the operators $\mathcal{L}_{IJ} $ still satisfy the algebra of $SO(2k+2)$.
\paragraph{Transverse Fluctuations}
The transverse fluctuations are those for which $\phi_j^{cl}=0$, and the gauge field components that are not in the normal direction to the defect. We will call $D$ the number of non-zero scalar matrices. For these excitations the only surviving mass term is of the form
\begin{equation}
\tr\bigg[[t_I, \delta\phi][t_I, \delta \phi]\bigg]= - \tr\bigg[\delta \phi[t_I,[t_I, \delta \phi]]\bigg]= -\tr\bigg[\delta \phi\,\sum_I\mathcal{L}^2_{I,\star}\, \delta\phi\bigg],
\end{equation}
which is the Laplacian on the fuzzy sphere, and the label $\star$ is $D+1$ for even dimensional spheres and $D+2$ for odd dimensional spheres. The eigenvalues can be determined by decomposing the color indices into irreducible representations of $SO(D)$. For even dimensional spheres the fuzzy Laplacian may be rewritten in terms of quadratic Casimirs or $SO(D+1)$ and $SO(D)$ \cite{Medina:2002pc}:

\begin{equation}
\frac{1}{2}\sum_{I=1}^D\mathcal{L}^2_{I,D+1}= \frac{1}{2}\sum_{1\leq I< J\leq D+1}\mathcal{L}^2_{I,J}-\frac{1}{2}\sum_{1\leq I< J\leq D}\mathcal{L}^2_{I,J}=C_2^{SO(D+1)}-C_2^{SO(D)}.
\end{equation}
For odd dimensional fuzzy spheres there is a subtlety since the matrices $\mathcal{L}_{I,J}$ require a projection into a subspace. The formula in that case is 
\begin{equation}
\frac{1}{2}\sum_{I=1}^{D}\hat{\mathcal{L}}^2_{I,D+2}=C_2^{SO(D+2)}-C_2^{SO(D+1)}-\frac{1}{2}\hat{\mathcal{L}}^2_{D+1,D+2}
\end{equation}
For the adjoint block, the last operator  $\hat{\mathcal{L}}^2_{D+1,D+2}$ acts non-trivially on the modes coming from $\text{Hom}(\mathcal{R}_\pm, \mathcal{R}_\mp)$ where it acts as a multiple of the identity with eigenvalue 4. For the off-diagonal mode $\hat{\mathcal{L}}^2_{D+1,D+2}$ acts as $L_{D+1, D+2}^2$ which is the identity matrix so its eigenvalue is 1. 
\paragraph{Longitudinal Fluctuations}
The mixing for the longitudinal fields is more complicated, mainly because both the color and flavor indices mix. Let us first consider the scalar fluctuations $\tilde{\phi}_I$ with $I=1, \dots, D$. Besides the Laplacian term there are now terms of the form 
\begin{equation}
\tr\bigg[ \frac{1}{2}[t_I, t_J][\tilde{\phi}_I, \tilde{\phi}_J]+ \frac{1}{2}[t_I, \tilde{\phi}_I][t_J,\tilde{\phi}_J ]+ \frac{1}{2}[t_I, \tilde{\phi}_J][\tilde{\phi}_I, t_J]\bigg].
\end{equation}
By using trace identities we can rewrite these as
\begin{equation}
-\tr\left[\tilde{\phi}_I [[t_I, t_J], \tilde{\phi}_J] \right]\simeq -\tr\left[\vec{\tilde{\phi}}^T\left(S_{IJ}\mathcal{L}_{IJ} \right)\vec{\tilde{\phi}}\right].
\end{equation}
Where  $S_{IJ}$ are $SO(D)$ generators in the fundamental representation and $\mathcal{L}_{IJ}$ are adjoint derivations with respect to $G_{IJ}$. The normal component of the gauge field mixes with the scalars, so we can assemble all the longitudinal modes into a $D+1$ vector $\vec{C}= (\tilde{\phi}_I, A_z)$. The mass matrix is given by\footnote{The factor of two difference in off-diagonal terms relative to \cite{Gimenez-Grau:2019fld} come from our choice of normalization for the $G_I$ matrices. This results in an overall factor of 4 in our masses relative to theirs.} 
\begin{equation}
\tr\bigg[C^T \begin{pmatrix}
\frac{1}{2}\sum_I\mathcal{L}_{I, \star}^2- \frac{1}{2}S_{IJ} \mathcal{L}_{IJ}&& 2i\sqrt{2}\, \hat{e}_I \mathcal{L}_{I, \star}\\ \\
-2i\sqrt{2}\,\hat{e}_I^T \mathcal{L}_{I, \star}&& \frac{1}{2}\sum_I\mathcal{L}_{I, \star}^2
\end{pmatrix} C \bigg],
\end{equation}
where $\hat{e}_I$ are $D$-dimensional unit vectors. The last step is to rewrite the interaction between color and flavor indices in terms of Casimir operators:
\begin{equation}
S_{IJ} \mathcal{L}_{IJ}=\sum_{1\leq I<J\leq D} J_{IJ}^2- \mathcal{L}_{IJ}^2- S_{IJ}^2.
\end{equation}
As before there is a slight difference in the cases of even dimensional and odd dimensional spheres. It turns out that once we diagonalize the mass matrix for a fuzzy $S^{2k}$ defect, the spectrum of fluctuations of a fuzzy $S^{2k-1}$ can be obtained easily, since the corresponding harmonics can be embedded in the harmonics of a sphere in one dimension higher. The only subtlety is that we have take into account the contributions to the mass coming from the $\hat{\mathcal{L}}^2_{D+1,D+2}$ term. We can always chose a basis for the generators of the algebra such that this generator is simultaneously diagonalizable with the remaining quantum numbers, so one needs to perform a further splitting of the $S^{2k}$ harmonics. 
\subparagraph{Fuzzy $S^3$}
Now we will consider the mass matrix for the fuzzy $S^3$ defect, since the details for the $S^4$ have been worked out in detail in \cite{Gimenez-Grau:2019fld}.

For the transverse scalar modes of the fuzzy $S^3$ we simply need to diagonalize the fuzzy Laplacian term. In the adjoint blocks its eigenvalues are given by \footnote{Our masses differ from those in  \cite{Gimenez-Grau:2019fld} since we chose to normalize the mass term such that it is of the form $\frac{1}{2} \delta \phi M^2 \delta\phi$.}
\begin{equation}
\begin{aligned}
m^2_{\pm \pm}(L_1, L_2)&=4\left(2L_1L_2 +L_2+ 2 L_2\right)\\
m^2_{\pm \mp}(L_1, L_2)&=4\left(2L_1L_2 +L_2+ 2 L_2-1\right),
\end{aligned}
\end{equation}
where the $\pm$ labels refer to blocks associated to the maps between the spaces $\mathcal{R}_\pm$, while the off-diagonal blocks have masses
\begin{equation}
m^2_{m,a'}= \left(n(n+4)-1\right)
\end{equation}
where $(L_1, L_2)$ take values according to the decompositions of $\text{End}(\mathcal{R}_{\pm}) $ and $\text{Hom}(\mathcal{R}_+, \mathcal{R}_-)$ and $\mathcal{R}_+\oplus \mathcal{R}_-$. Note that the weights of $SO(4)$ are related to the $SU(2)\times SU(2)$ spins by $(L_1+L_2, L_1-L_2)$.
Formally these modes are expanded in terms of fuzzy spherical harmonics $\hat{Y}_{(\boldsymbol{L_1}, \boldsymbol{L_2})}$ \cite{Ramgoolam:2001zx} which form a basis for the states in tensor decompositions mentioned above.

For the longitudinal scalar modes of the fuzzy $S^3$, the relevant $SO(4)$ representation for $S_{IJ}$ is $(\frac{1}{2}, \frac{1}{2})$. First we need diagonalize the total quadratic Casimir operator $J^2$. This amounts to decomposing the tensor product of the fundamental representation of $SO(4)$ with $(L_1, L_2)$.
\begin{equation}
\begin{aligned}
&\left(\frac{1}{2}, \frac{1}{2}\right)\otimes (L_1, L_2)\cong \\
&\left(L_1+\frac{1}{2}, L_2+\frac{1}{2}\right)\oplus \left(L_1-\frac{1}{2}, L_2+\frac{1}{2}\right)\oplus  \left(L_1+\frac{1}{2}, L_2-\frac{1}{2}\right)\oplus  \left(L_1-\frac{1}{2}, L_2-\frac{1}{2}\right)
\end{aligned}
\end{equation}
This diagonalizes a $4\times 4$ sub-block of the complicated mass matrix, but there is still mixing with the normal component of the gauge field. The new harmonics resulting from the tensor product with the fundamental representation transform as bi-spinors of $SO(3)\times SO(3)$. More explicitly  the scalar fluctuations assemble into a bi-spinor field

\begin{equation}
\vec{B}_{\boldsymbol{L}}^{\alpha \beta} C^{\boldsymbol{J}}_{\boldsymbol{L}, \alpha \beta} \;\mathcal{Y}_{\boldsymbol{J}}
\end{equation}
where $\bold{J}=(J_{1},J_{2} )=(L_{1} \pm\frac{1}{2}, L_{2} +\beta \frac{1}{2})$, and $\alpha, \beta= \pm 1$, and $C^{\boldsymbol{J}}_{\boldsymbol{L}, \alpha \beta} $ is a Clebsh-Gordan coefficient. These bi-spinor harmonics can be rewritten in terms of ordinary fuzzy harmonics with shifted indices.
The off-diagonal terms of the mass matrix can be written in terms of tensor operators.

To evaluate the matrix elements of the off-diagonal part of the mass matrix we need embed the $SO(4)$ basis into an $SO(5)$ representation. Using the notation of \cite{Caprio:2010tj}, an $SO(5)$ irreducible representation is labeled by a set of weights $(R,S)$ with $R\geq S \geq 0$ and the states are indexed by a set of $SO(3)\times SO(3)$ quantum numbers $(X,Y)$. The labels are related by 
\begin{equation}\label{so(4) restriction}
\begin{aligned}
X&= R-\frac{1}{2}n - \frac{1}{2}m\\
Y&= S+\frac{1}{2}n-\frac{1}{2}m\\
0&\leq n\leq 2(R-S)\\
0&\leq m\leq 2S.
\end{aligned}
\end{equation}
The symmetric tensor representation of $SO(5)$ is labeled by $(\frac{n}{2},0)$. The adjoint block then transforms in the 
\begin{equation}
\left(\frac{n}{2},0\right)\otimes \overline{\left(\frac{n}{2},0\right)}= \bigoplus\;(R,S),
\end{equation}
with $0\leq S\leq R$ and $R+S\leq n$. Within each the $(R,S)$ summands we have to further project into the appropriate $\mathcal{R}\otimes \bar{\mathcal{R}}$; this comes from the projection from the fuzzy $S^4$ into the fuzzy $S^3$. In practice we can do this by decomposing the appropriate tensor products of $SO(3)\times SO(3)$ irreducible representations according to the restriction \ref{so(4) restriction}.  This means that the modes on the adjoint blocks carry labels $(R,S: L_1,L_2)$, where $(R,S)$ label the $SO(5)$ quantum numbers and $(L_1,L_2)$ the $SO(4)$ quantum numbers with the ranges given by the decomposition of $\mathcal{R}\otimes \bar{\mathcal{R}}$. The mixing with the gauge field involves the evaluation of matrix elements of the form
\begin{equation}
\langle J_1, J_2; m_{J_1}, m_{J_2}|T^{(\frac{1}{2}, \frac{1}{2}), \frac{1}{2}, \frac{1}{2}}_{\alpha \beta}| L_1, L_2; m_{L_1}, m_{L_2}  \rangle.
\end{equation}
Note that the spins $(J_1, J_2)$ implicitly depend on $\alpha, \beta$. The spherical tensors $T^{(\frac{1}{2}, \frac{1}{2}), \frac{1}{2}, \frac{1}{2}}_{\alpha \beta}$ are related to $SO(6)$ generators by a change of basis;
\begin{equation}
\hat{e}_I L_{I6}= \hat{c}_{\alpha \beta} T^{(\frac{1}{2}, \frac{1}{2}), \frac{1}{2}, \frac{1}{2}}_{\alpha \beta},
\end{equation}
where $ \hat{c}_{\alpha \beta}$ are basis vectors for $\left(\frac{1}{2}, \frac{1}{2} \right)$ representation of $SO(5)$  restricted to the four dimensional bi-spinor of $SO(4)$. 
 The Wigner-Eckard theorem implies that this matrix element factorizes into am $SO(4)$ Clebsch-Gordan coefficient and a reduced matrix element; the $SO(4)$ Clebsch-Gordan coefficients are products of $SO(3)$ coefficients and the reduced matrix elements were computed in \cite{HECHT1965177, Caprio:2010tj}. The diagonalization then proceeds as with the case of the fuzzy $S^4$ and we reproduce the details in the appendix.

For the fermion mass matrix we can diagonalize $\mathcal{C}^\dagger \mathcal{C}$, we can borrow the results for the fuzzy $S^4$.
\begin{equation}
\mathcal{C}^\dagger \mathcal{C}= \frac{1}{2}\sum_{I=1}^4 L_{I,5}^2- \tilde{S}_{IJ}L_{IJ},
\end{equation}
where the $\tilde{S}_{IJ}$ are four dimensional generators of $SO(4)$. The fermions will transform in two copies (one for each chirality) of the reducible representation $(\frac{1}{2},0)\oplus (0, \frac{1}{2})$ which restricts from the six dimensional spinor representation. This is the same problem as determining the eigenvalues of the $4\times 4$ block for the complicated bosons, except that the allowed values of $(J_1,J_2)$ are $(L_1 \pm \frac{1}{2}, L_2)$ and $(L_1, L_2 \pm \frac{1}{2})$. 

\section{Propagators}

\paragraph{Defect CFT Kinematics}
To keep the discussion as general as possible we will work in Euclidean signature in dimension $d+1$, and will work with conformal defects along a $\mathbb{R}^{p+1}$ subspace. This is useful for performing perturbative calculations using dimensional regularization. The metric can be decomposed into $p+1$ coordinates $t, x_1, \dots, x_p$ along the defect, and $d-p$ coordinates transverse coordinates $y_\alpha$:
\begin{equation}
ds^2= dt^2 + \sum_{i=1}^p dx_i^2 + \sum_{\alpha=1}^{d-p} dy_\alpha^2.
\end{equation}
This configuration breaks the conformal symmetry into $SO(1,p+1)\times SO(d-p)$, where $SO(1,p+1)$ is the conformal group along the defect and $SO(d-p)$ is the rotational symmetry of the directions transverse to the defect. Another useful coordinate choice comes from placing the defect at $y_\alpha=0$ and using the transverse distance to the defect as a radial coordinate $z= \sum_{\alpha}y_\alpha y_{\alpha}$. The resulting metric is conformally equivalent to $EAdS_{p+2}\times S^{d-p-1}$ in Poincare coordinates
\begin{equation}\label{FlattoAdS}
ds^2= z^2\left(\frac{dt^2+ dz^2+\sum_{i=1}^p dx_i^2 }{z^2}+ d\Omega_{d-p-1}^2\right).
\end{equation}
In these coordinates the flat space Green's function in the background of the defect is related to solving for the Green's function of a massive field in $AdS$. This mass is determined by diagonalizing the quadratic part of the action around the classical configuration associated to the defect and adding an additional mass term associated to the angular momentum on $S^{d-p-1}$. The kinematic part of the propagators is fixed by the residual conformal symmetry of the system. For completeness we will list the propagators for various kinds of fields. 

\paragraph{Scalar Propagator}
The quadratic fluctuations of a scalar field around the defect have equations of motion of the form
\begin{equation}
\left(-\partial_\mu \partial^\mu + \frac{m^2}{z^2} \right)\delta\phi(t, x_i, y_\alpha)=0,
\end{equation}
where the index contractions are performed with respect to the flat metric, and $z= y_{\alpha}y_{\alpha}$. The propagator solves the following equations
\begin{equation}
\left(-\left[\partial_t^2 +\partial_{x_i}^2+ \partial_z^2 + \frac{(d-p-1)}{z}\partial_z + \frac{1}{z^2} \Delta^2_{S^{d-p-1}}\right] + \frac{m^2}{z^2} \right)K(x,x')=\delta(x,x').
\end{equation}
In $AdS\times S$ coordinates, rescale the propagator $K= \frac{1}{(zz')^{(d-1)/2}} \tilde{K}$ which lets us rewrite the equation as a the Green's function for a massive scalar field on $AdS_{p+2}\times S^{d-p-1}$. 

\begin{equation}
\begin{aligned}
\left[-\Delta_{AdS_{p+2}}- \Delta_{S^{d-p-1}} +\tilde{m}^2\right]\tilde{K}(x,x')&= z^{d+1} \delta(x-x'),\\
\tilde{m}^2 &= m^2- \left(\frac{d-1}{2}\right)\left(\left(\frac{d-1}{2}\right)-(p+1) \right).
\end{aligned}
\end{equation}
The $z^{d+1}$ is the Weyl factor for the conformal transformation \eqref{FlattoAdS}, so it should be reabsorbed into the definition of the delta function on $AdS_{p+2}\times S^{d-p-1}$. The shift in the mass term has a natural explanation. A conformally coupled scalar in $AdS_{n}\times S^{m}$ has a mass term due to coupling with the curvature
\begin{equation}
\xi R_{AdS_{n}\times S^{m}} = \frac{(n+m-2)}{4(n+m-1)}\left( \frac{m(m-1)}{L_S^2}-\frac{n(n-1)}{L_{AdS}^2} \right).
\end{equation}
In our case the conformal transformation that takes flat space into $AdS_{p+2}\times S^{d-p-1}$ has $L_{AdS}=L_S$ which we can set to one by a rescaling. The shift in the mass due to the rescaling of the propagator is precisely the conformal coupling of the scalar to the background. In other words we could have started with the theory on  $AdS_{p+2}\times S^{d-p-1}$  where the boundary is associated to the position of the defect.
 We can further simplify this by restricting to fields that transform in irreducible representations of $SO(d-p)$. These would correspond to either spinning primaries and descendant states. This is completely equivalent to starting with the theory on $AdS_{p+2}\times S^{d-p-1}$ and performing a KK reduction on the sphere by expanding the fields in harmonics. The effective masses of the scalars are $m_{L,d,p}^2= \tilde{m}^2+ L(L+d-p-2)$ and we are left with the task of computing the Green's function for a tower of massive scalars in $AdS_{p+2}$:
\begin{equation}
K_{\text{AdS}}^{\nu}(x,x')= (z z')^{\frac{p+1}{2}} \int \frac{d^{p+1}\vec{k}}{(2\pi)^{p+1}} e^{i \vec{k}\cdot (\vec{x}-\vec{x}')} I_\nu (|\vec{k}| z^{<}) K_\nu(|\vec{k}| z^>),
\end{equation}
with $z^{<}$ being the minimum between $z$ and $z'$ and similarly for $z^>$.
\paragraph{Spinor Propagator}
The propagator for the spinor fields can be obtained from the scalar propagators. Just as with the scalar we can start with the metric conformally rescaled to $AdS\times S$, so the Dirac operator splits into two terms.
\begin{equation}
i \gamma^\mu D_\mu = i \slashed{D}_{\text{AdS}} + i \slashed{D}_{S}
\end{equation} After that we expand the propagator into spinor harmonics. 
\begin{equation}
K_F(x,x', \Omega, \Omega') = \sum_{\bold{J}} K_F^{\nu(m)}(x,x') \,\mathcal{Y}_{\boldsymbol{J}}^* (\Omega)\mathcal{Y}_{\boldsymbol{J}}(\Omega'),
\end{equation}
note that $\nu$ implicitly depends on the spin quantum numbers $\boldsymbol{J}$. The fact that $\mathcal{Y}_{\boldsymbol{J}}(\Omega)$ are spinor harmonics modifies the mass of the spinor relative to mass of a scalar field.
Then we are left with the problem of finding the Green's function for a massive spinor in $AdS$, which we can obtain from the scalar propagator in a standard way. 
\begin{equation}
K_{F}^{\nu}(x,x')= \sqrt{\frac{z'}{z}}\left[i \slashed{D}_{\text{AdS}}  +\frac{i}{2} \gamma^{\star} +m \right]\left[ K_{\text{AdS}}^{\nu(m-\frac{1}{2})}(x,x')\mathcal{P}_- +K_{\text{AdS}}^{\nu(m+\frac{1}{2})}(x,x')\mathcal{P}_+\right],
\end{equation}
with $\gamma^\star$ standing for the gamma matrix associated with the normal direction to the defect, and $\mathcal{P}_\pm \frac{1}{2}(1\pm i \gamma^\star) $.

\paragraph{Codimension 3 case: Monopole Harmonics}
The codimension $3$ case is special because the geometry is $AdS_2\times S^2$, and it is possible to have non-trivial electromagnetic flux threading the $S^2$. This is taken care of by expanding the modes on $S^2$ in terms of monopole harmonics. In that case we need to add an additional gauge fixing term 
\begin{equation}
\partial_{\mu} \tilde{A}^\mu +i[\tilde{A}^\mu, A^{cl}_\mu]+ i[\tilde{\phi}_i, \phi_i^{cl}]=0.
\end{equation}
In this case there are two different types of modes, those which sit in a diagonal block, or those that sit in an off-diagonal block. For simplicity we will restrict to the case where only one of the eigenvalues of the gauge field gets a non-trivial profile, but the analysis extends to general monopole configurations. Fluctuations of fields within a diagonal mode are uncharged and they do not feel the monopole background; for these the propagators are as discussed in the previous sections. Off-diagonal modes are charged and transform as sections of a non-trivial line bundle over the  $S^2$ transverse to the defect. In this case there is mixing between the gauge field components normal to the defect and the scalar fields which do not carry a vev; like before we call these complicated fields. The remaining fields do not mix and we call them simple fields. This analysis was carried over in \cite{Kristjansen:2023ysz}, see also \cite{Wu:1976qk, Olsen:1990jm, Weinberg:1993sg}.
The mode expansion for the easy fields is done in terms of monopole harmonics :

\begin{equation}
\begin{aligned}
\phi_{\text{easy}}&= \sum_{l, \bar{m}} \phi^{l, \bar{m}}(z,t) \,Y^{(q)}_{l, \bar{m}}(\theta, \phi)\\
Y^{(q)}_{l\bar{m}}&= C_{l\bar{m}}\left(1-\cos\theta\right)^{\alpha/2} \left(1+\cos \theta \right)^{\beta/2} P^{(\alpha, \beta)}_{l-\frac{\alpha}{2}-\frac{\beta}{2}}(\cos \theta) \,e^{i(\bar{m}-\frac{B}{2}) \phi}.
\end{aligned}
\end{equation}
where $q=B/2$ is the monopole charge, $\bar{m}= m-\frac{B}{2}$ is the eigenvalue of $L_z$, and $C_{l\bar{m}}$ is a normalization constant, and $P^{(\alpha, \beta)}_{l}$ are Jacobi polynomials. The parameters $\alpha$, $\beta$ are related to the monopole charge and angular momentum quantum number by
\begin{equation}
\begin{aligned}
\alpha &= |\bar{m}+ \frac{B}{2}|\\
\beta &= | \bar{m}-\frac{B}{2}|.
\end{aligned}
\end{equation}
The monopole harmonics are eigenfunctions of the Laplace operator $D_{q}^2= (\nabla_\mu+i A^{cl}_\mu)^2$ with eigenvalues $l(l+1)$, where $l\geq |q|$, and $\bar{m}= -l, -(l-1), \dots, l-1, l$. The remainder of the analysis proceeds as the other cases.

For the complicated modes one needs to expand in vector monopole harmonics and diagonalize the kinetic term matrix. The vector monopole harmonics are constructed by taking a tensor product with a $s=1$ SU(2) representation and decomposing into irreducible representations. They are eigenfuctions of the total angular momentum operator $(\vec{L}+\vec{S})^2$ and of $L^2$ and $S^2$
\begin{equation}
\begin{aligned}
(L+S)^2 \boldsymbol{Y}^{(q)}_{J lM}(\theta, \phi)&= J(J+1) \boldsymbol{Y}^{(q)}_{J lM}(\theta, \phi)\\
(L_z+S_z) \boldsymbol{Y}^{(q)}_{J lm}(\theta, \phi)&= M\boldsymbol{Y}^{(q)}_{J lM}(\theta, \phi)\\
L^2 \boldsymbol{Y}^{(q)}_{J lm}(\theta, \phi)&= l(l+1) \boldsymbol{Y}^{(q)}_{J lM}(\theta, \phi)\\
S^2 \boldsymbol{Y}^{(q)}_{J lm}(\theta, \phi)&= 2 \,\boldsymbol{Y}^{(q)}_{J lM}(\theta, \phi).
\end{aligned}
\end{equation}
They also satisfy the following product formulas:
\begin{equation}
\begin{aligned}
\boldsymbol{\hat{r}} \cdot \boldsymbol{Y}^{(q)}_{J lM}(\theta, \phi)&= \mathcal{C}^{(q)}_{Jl} Y^{(q)}_{JM}(\theta, \phi)\\
\boldsymbol{\hat{r}} \times \boldsymbol{Y}^{(q)}_{J lm}(\theta, \phi)&=i \sum_{L} \mathcal{A}^{(q)}_{Jl L}\boldsymbol{Y}^{(q)}_{J LM}(\theta, \phi),
\end{aligned}
\end{equation}
where $\mathcal{C}$ and $\mathcal{A}$ are Clebsch-Gordan coefficients \cite{Weinberg:1993sg}. The size of the kinetic term matrix depends on the number of scalar fields that get a non-trivial profile. The supersymmetric case was considered in detail in\cite{Kristjansen:2023ysz}.

\section{Holographic interpretations}

\paragraph{Codimension 1}
The integrable codimension one defects for the fuzzy two and four sphere correspond to D5 and D7 probe brane set-ups. These have been studied at both quantum and classical level and are by now quite well-understood.  

\paragraph{Codimension 2}
The simplest example in codimension two is the Gukov-Witten defect \cite{Gukov:2006jk}. These are realized holographically by a probe $D3$ brane wrapping an $AdS_3\times S^1 \subset AdS_5\times S^5$. The defect preserves a $\mathfrak{psu}(1,1|2)^2\rtimes \mathfrak{su}(2)_R $ subalgebra of the superconformal algebra $\mathfrak{psu}(2,2|4)$. If we insist on having classical solutions that manifest conformal symmetry the only allowed defects have diagonal scalar fields, but additional solutions with noncommuting matrices do exist although they have less singular behavior \cite{Gukov:2008sn}. This defect appears in the classification of half-BPS integrable boundary conditions for the sigma model at strong coupling \cite{Dekel:2011ja}. In the classification of half supersymmetric codimension two defects there is also a defect preserving a $\mathfrak{su}(1,1|4)\times \mathfrak{su}(1,1)$ subalgebra which has a natural interpretation as a $D7$ brane wrapping an $AdS_3\times S^5$ \cite{Harvey:2008zz}. One would expect that this brane needs to stabilized by flux as in the codimension one case D3-D7 defect set up, which would lead to a noncommutative $S^5$. We can identify this with the fuzzy $S^5$ solution attached to a codimension two defect. It would be instructive to check the supersymmetry of this defect. This exhausts all the possible half-BPS codimension two defects.
For non-supersymmetric defects there is a codimension two fuzzy $S^3$. This may be identified with a flux stabilized fivebrane wrapping an $AdS_3\times S^3$. This would be similar to the non-supersymmetric $AdS_4\times S^4$ D7 defect in codimension one. The remaining possibilities cannot be realized by supersymmetric D-branes of the type IIB string. These cases are the fuzzy $S^4$ and fuzzy $S^2$ without gauge field singularities. Because these fuzzy spheres arise as the equators of the other cases it is possible that these defects are either unstable nonsupersymmetric branes \cite{Sen:1999mg}, or restrictions coming from intersections of the other cases. 

\paragraph{Codimension 3}
The codimension $3$ case requires a bit more care. Again we start with half-BPS configurations; the only possibility are defects preserving $\mathfrak{osp}(4^*|4)$. These are either a D1/F1 defect wrapping $AdS_2$, D3 brane wrapping $AdS_2\times S^2 \subset AdS_5$ or a D5 brane wrapping $AdS_2 \times S^4$. Only the D1 and D5 defects are known be integrable at strong coupling. The D1/F1 set up is related to fundamental Wilson loops or to 't Hooft loops. These correspond to monopole solutions with a single diagonal scalar field; their weak coupling integrability was studied in \cite{Kristjansen:2023ysz}. The only remaining half-supersymmetric integrable defect is the $AdS_2\times S^4$ D5 brane. This is described by a Wilson loop in the fully antisymmetric representation. We do not expect that there exists a classical 't Hooft loop description of this defect, and instead its description should be closer to that of a maximal D3 giant graviton. The reason one expects this is that fivebrane configurations of matrix models are usually associated to the trivial vacuum at strong coupling \cite{Asano:2017nxw}. 

More generally, the equations of motion for the scalars in codimension three force the profiles to be purely diagonal matrices. The gauge constraints tie the vev of the fields to the electric and magnetic charges, so no fuzzy sphere classical solutions are possible in codimension 3.

\section{Conclusions}
We studied matrix product states associated to classical solutions of $\mathcal{N}=4$ SYM with pole singularities along surfaces of arbitrary codimension. These are associated to weakly coupled descriptions of conformal defects of the model. By studying the action of one of the odd charges of the model, $Q_3$, we found that all integrable matrix product states of fixed bond dimension are necessarily described by fuzzy spheres, which we argue extends to any $SO(n)$ Heisenberg chain. We provided evidence for the integrability of the remaining fuzzy sphere MPS. This gives a partial classification of integrable MPS for this family of integrable models. 

A natural conjecture is that the action of $Q_3$ might be sufficient for testing the integrability of a given MPS. Certainly we expect to be true for $SO(n)$ chains. A plausible explanation is that $Q_3$ on an MPS is closely related to matrix equations of the type
\begin{align}
\sum_I [t_I, [t_I, t_k]]= R^2 t_k,
\end{align}
which are noncommutative analogs of the equations for a harmonic map. Roughly speaking one can think of the $t_k$ as certain coordinates on an extremal hypersurface on an $n+1$ dimensional sphere. These are naturally described by embedding maximal spheres of lower dimension in $S^{n+1}$. This is very similar to the expectation coming from string theory, where the integrable boundaries are usually associated to cycles of maximal size. It would be interesting to make these ideas precise. 

We do not expect the classification given here to be exhaustive. One reason is that there are additional boundary conditions coming from intersecting branes that are expected to be integrable. For instance the case of maximal giant gravitons intersecting at angles comes to mind \cite{Bajnok:2013wsa, Zhang:2015fea, Holguin:2021qes}. Operators describing these configurations preserve one-quarter of the supersymmetries and their explicit form is not known. Very near the intersection, the branes can be taken to look like in flat space and the structure of the boundary states for these configurations is already very non-trivial in flat space. It would be nice to understand the spin chain analogs of these configurations, particularly the notion of a supersymmetric boundary condition for the spin chain. The analogous structure in conformal field theory is that of a matrix factorization category. It would also be interesting to study defect configurations that lead to non-trivial gauge field profiles. The case of the 't-Hooft monopole was studied in detail in, but a similar analysis for Gukov-Witten surface defects is still lacking. The one-point functions of protected operators have been studied in those set ups in the semiclassical approximation\cite{Drukker:2008wr} and using supersymmetric localization \cite{Choi:2024ktc}, but correlation functions of non-protected operators remain to be studied. These defects are expected to be integrable by strong coupling considerations and by the arguments presented here. It would be particularly interesting to study the singular limits of such defects which lead to non-diagonal vevs for both scalars and gauge fields \cite{Gukov:2008sn}.

Another direction that needs attention is the classification of crosscap states for integrable chains \cite{Caetano:2021dbh}. In this case the non-local nature of the entanglement structure of the boundary state makes the arguments used here cumbersome. By extrapolating ideas from the continuum string description one expects that there exist more general crosscap-type states associated with gauging of the parity along the chain. These configurations should be viewed as spin chain analogs of orientifold states in 2d CFT \cite{Dai:1989ua}. The relation of these spin chain states and defects in $\mathcal{N}=4$ also needs further clarification \cite{Caetano:2022mus}. 

A next step would be to extend and prove the overlap formulas for the fuzzy $S^3$ and fuzzy $S^5$ to the full $SO(6)$ sector. The techniques developed in \cite{Gombor:2021uxz, Gombor:2021hmj, Gombor:2022deb, Gombor:2023bez} could be useful for determining the overlaps for bond dimension as well as for checking the integrability of the higher dimensional fuzzy sphere matrix product states in general $SO(m)$ Heisenberg chains. This information is important for implementing the defect CFT bootstrap program in $N=4$ SYM \cite{Caron-Huot:2017vep, Liendo:2019jpu, Lemos:2017vnx}, for instance in codimension two. We expect that the codimension-2 fuzzy $S^5$ defect preserves half of the supersymmetries \cite{Wang:2020seq}, so it should be associated to a stable and integrable defect CFT. One reason to consider this case is that the inversion formulas can be applied in a more straightforward fashion in codimension-2 as opposed to the codimension-1 case. For this reason it would also be necessary to compute the leading order two point functions of chiral primaries in this set-up. In particular two-point functions of the stress-energy tensor are of interest and they have already be computed in the codimension one case \cite{deLeeuw:2023wjq}.

\paragraph{Acknowledgements} 
We would like to thank Daire Connolly for work in the early stages of this project. We are grateful to Tamás Gombor for discussions regarding overlap formulas for integrable matrix product states, and for verifying our formulas. We would also like to thank Charlotte Kristjansen for discussions and comments on the draft.
MdL was supported in part by SFI and the Royal Society for funding under grants UF160578, RGF$\backslash$ R1$\backslash$ 181011, RGF$\backslash$8EA$\backslash$180167 and RF$\backslash$ ERE$\backslash$ 210373. MdL and AH are also supported by ERC-2022-CoG - FAIM 101088193. 

\appendix

\section{Representation theory of $SO(n)$}
In this appendix we review some details about the representation theory of $SO(2k)$ and $SO(2k+1)$ that are necessary for dealing with higher dimensional fuzzy spheres. 
\subsection{$SO(2k+1)$}
Irreducible tensor representations of $SO(2k+1)$ can be put into one-to-one correspondence with Young diagrams of shape $\mu= (\mu_1, \mu_2, \dots, \mu_k)$. More precisely, from a rank $n$ tensor product of the defining vector representation $V$ one can build irreducible representations as follows. First one projects the space of rank $n$ tensors into a subspace of rank $n$ traceless tensors. This amounts to demanding that the representation is annihilated by the contraction (or trace) operation $K$. The rest of the construction is identical to the case of tensor representation for the unitary group $U(k)$, in that the remaining space is decomposed into tensor structures with fixes symmetry structure. To a partition $\mu= (\mu_1, \mu_2, \dots, \mu_k)$ of $n$ we can associate a Young diagram with some amount of rows $R$ and columns $C$. The number of boxes in the $i^{th}$ row is $\mu_i$ and we call the number of boxes in the $j^{th}$ column $c_j$. An irreducible tensor is then obtained by applying a Young symmetrizer. The group $S_\bold{R}= S_{\mu_1}\times S_{\mu_2}\times \dots \times S_{\mu_R}$ acts on the tensor by permuting the row indices, while $S_\bold{C}= S_{c_1}\times S_{c_2}\times \dots \times S_{c_C}$ permutes the  column indices of the tensor. The Young symmetrizer for the label $\mu$ is given by 
\begin{equation}
\mathcal{Y}_\mu=\frac{1}{|S_\bold{R}| |S_\bold{C}|}\sum_{\sigma \in S_\bold{R}}\sum_{\tau \in S_\bold{C}} (-1)^\tau \tau \sigma.
\end{equation}
The dimension of the irreducible representation associated to a Young diagram $\mu$ is given by 
\begin{equation}
\begin{aligned}
D^{SO(2k+1)}(\mathcal{R}_\mu)&= \prod_{i<j} \frac{(l_i^2-l_j^2)}{(m_i^2-m_j^2)}\,\prod_i \frac{l_i}{m_i}\\
l_i&= \mu_i+k-i+\frac{1}{2}\\
m_i&= k-i+\frac{1}{2}.
\end{aligned}
\end{equation}
In addition to the tensor representations, there exist spinor representations with half-integer labels $\mu_i \in \mathbb{Z}+\frac{1}{2}$:
\begin{equation}
\mu_1\geq \mu_2\geq \dots \geq \mu_k\geq \frac{1}{2}.
\end{equation}
The dimensions for the spinor representations are obtained using the same formula as for tensor representations, but with half-integer labels. The fundamental spinor representation is $(\frac{1}{2}, \frac{1}{2}, \dots, \frac{1}{2})$, and its $n$-fold symmetric tensor product has weights $(\frac{n}{2}, \frac{n}{2}, \dots, \frac{n}{2})$.

We also need the eigenvalues of the quadratic Casimir operator to compute the radii of fuzzy spheres and the masses of fluctuations around the defect. In terms of the weights $\mu_i$ this is given by 
\begin{equation}
C^{SO(2k+1)}_2(\mathcal{R}_\mu)= \sum_{i=1}^k 2\mu_i\left(\mu_i +2k+1-2i \right).
\end{equation}

\subsection{$SO(2k)$}
The representations for even special orthogonal groups are slightly more complicated. The tensor representations are labeled by $k$ integer weights $\mu_i$, except that the last label $\mu_k$ can be either positive, negative, or zero. If $|\mu_k|\neq 0$ there will exist two conjugate representations.  Tensor representations with $\mu_k>0$ are self-dual while those with $\mu_k<0$ are anti-self-dual.The construction of irreducible representations proceeds as in the case of odd special orthogonal groups if $\mu_k=0$, and whenever $|\mu_k|\neq0$ we add additional projectors into positive chirality $P_+= \frac{1}{2}\left(1+ \Gamma^{2k+1} \right)$ to each column of height $k$ of the Young diagram, and similarly for $\mu_k<0$ with $P_+$ replaced with $P_-$. The labels for spinor representations work in the same way: $\mu_i$ are half-integer with $\mu_k$ being allowed to take either positive or negative values. Now there are two fundamental spinor representations (one for each chirality) with weights $(\frac{1}{2}, \frac{1}{2}, \dots, \pm\frac{1}{2})$
\begin{equation}
\begin{aligned}
D^{SO(2k)}(\mathcal{R}_\mu)&= \prod_{i<j} \frac{(l_i^2-l_j^2)}{(m_i^2-m_j^2)}\\
l_i&= \mu_i+k-i\\
m_i&= k-i.
\end{aligned}
\end{equation} 
The eigenvalues of the quadratic Casimir in this cases are
\begin{equation}
C^{SO(2k)}_2(\mathcal{R}_\mu)= \sum_{i=1}^k 2\mu_i\left(\mu_i +2k-2i \right).
\end{equation}
\section{Details of mass matrix diagonalization}
\subsection{Matrix Elements of $SO(6)$ generators}
In the diagonalization of the longitudinal models for the fuzzy $S^3$ defect we are tasked to evaluate traces  of $\hat{G}_I$ matrices with fuzzy harmonics. This is equivalent to evaluating certain matrix elements of spherical tensors of $SO(4)$. To see why this is the case we can start by dropping the projectors and considering the matrices $G_I$ for $i=1,\dots 4$. These are not $SO(5)$ generators, but they can be viewed as generators of $SO(6)$ via $L_{I6}=G_I$.  We can view these as spherical tensors of $SO(5)$ and their matrix elements have been computed in \cite{Gimenez-Grau:2019fld}. The only additional complication that we need to deal with is the branching from $SO(5)$ into $SO(4)$ being careful to keep only the representation after applying the projection into the reducible representation into $\mathcal{R}$ . 
We consider the adjoint block fluctuations. We start by rewriting the scalar fields in a way that manifests $SO(3)\times SO(3)$ quantum numbers
\begin{equation}
\begin{aligned}
C_{++}&= \frac{-i}{\sqrt{2}} (\delta \phi_1+ \delta \phi_4)\\
C_{+-}&=\frac{1}{\sqrt{2}}(\delta \phi_1- \delta \phi_4)\\
C_{-+}&=\frac{-i}{\sqrt{2}}(\delta \phi_2- \delta \phi_3)\\
C_{--}&=\frac{1}{\sqrt{2}}(\delta \phi_2+ \delta \phi_3)\,
\end{aligned}
\end{equation}
The $C_{\alpha \beta}$ are bispinors of $SO(3)\times SO(3)$ transforming in the $(\frac{1}{2}, \frac{1}{2})$  representation, but it will be convenient to keep in mind that they transform in the five dimensional  $(\frac{1}{2}, \frac{1}{2})$ representation of $SO(5)$. The off-diagonal parts of the mass matrix involve tensors of the form

\begin{equation}
\hat{e}_I L_{I6}= \hat{c}_{\alpha \beta} T^{(\frac{1}{2}, \frac{1}{2}), \frac{1}{2}, \frac{1}{2}}_{\alpha \beta},
\end{equation}
where $T^{(\frac{1}{2}, \frac{1}{2}), \frac{1}{2}, \frac{1}{2}}_{\alpha \beta}$ is an $SO(5)$ spherical tensor and we rewrote the four dimensional representation of $SO(4)$ in terms of $SO(5)$ highest weight quantum numbers. Unlike the case of the fuzzy $S^4$ the labels $\alpha, \beta$ only take the values $\pm \frac{1}{2}$. At this point we can expand the fluctuations into $SO(4)$ fuzzy harmonics. We can formally write these as
\begin{equation}
C_{\alpha \beta}= \sum_{\bold{L}} C_{\alpha \beta}^{\bold{L}}\,|\bold{L}\rangle=\sum_{\bold{J}} B_{\alpha \beta}^{\bold{J}}\,|\bold{J}\rangle,
\end{equation}
where $|\bold{L}\rangle$ are $SO(5)$ basis vectors restricted to the appropriate $SO(4)$ subspaces and in the last equality we decomposed the tensor product of the  $(\frac{1}{2}, \frac{1}{2})$ representation with $(L_1, L_2)$. The off-diagonal parts of the mass matrix involve matrix elements of the form

\begin{equation}
\langle \bold{J}|T^{(\frac{1}{2}, \frac{1}{2}), \frac{1}{2}, \frac{1}{2}}_{\alpha \beta}| \bold{L}\rangle,
\end{equation}
and since we can use $SO(5)$ labels for the $SO(4)$ representations we find that the non-trivial part of off-diagonal terms of the mass matrix for the $S^3$ are identical to those of the fuzzy $S^4$ with the diagonal entries replaced accordingly. The only non-zero matrix elements are those with $(\alpha, \beta)= (\pm \frac{1}{2}, \mp\frac{1}{2}
)$
\begin{equation}
\begin{aligned}
&\sum_{I=1}^4\hat{e}_I L_{I6} \big|\left(L_1\pm \frac{1}{2},L_2 \mp \frac{1}{2}\right)\big\rangle= T^{\pm \mp} |(L_1, L_2)\rangle\\
 T^{+-} 
 &=  \sqrt{2} \sqrt{
    \frac{\left(2 J_1+1\right) \left(J_1-J_2\right) \left(J_2+1\right)}{2 J_1-2 J_2+1}
 }\\
 T^{-+}
 &= -\sqrt{2} \sqrt{
    \frac{\left(2 J_1+3\right) \left(J_1-J_2+1\right) J_2}{2 J_1-2 J_2+1}
 }.
\end{aligned}
\end{equation}
From here the details are virtually identical to the fuzzy $S^4$ case, except that the allowed values of $(L_1, L_2)$ are taken from \eqref{dimensions} and the quadratic Casimirs are taken with respect to $SO(4)$. Note that there is mixing between the $2\times 2 $ blocks in the decomposition of the tensor product $\mathcal{R}\otimes \bar{\mathcal{R}}$. In $SO(5)$ language this has to do with the fact that the matrix elements of the generators $L_{I6}$ are non-zero for $(J_1, J_2)= (L_1 \pm \frac{1}{2}, L_2 \mp\frac{1}{2})$ which mixes the blocks within themselves. The non-trivial part of the mass matrix is

\begin{equation}
        \begin{pmatrix}
           -\frac{1}{2}\left(J^2-L^2-S^2+L_{56}^2 \right) & -i\sqrt{2}/,4 \hat{c}_{\alpha \beta}T^{\alpha \beta} \\
          \\
          i \sqrt{2} /,2\hat{c}_{\alpha \beta}^\dagger T^{\alpha \beta} &-\frac{1}{2}L_{56}^2.
        \end{pmatrix}
\end{equation}
With our choice of basis $L_{56}$ is a Cartan generator of $SO(6)$, so it can be simultaneously diagonalized with the remaining $SO(4)$ generators.The eigenvectors are given by modes $B_{\alpha \beta}$. The idea is then to reduce the fields with $SO(5)$ labels $B_{\alpha \beta}$ into fields with $SO(4)$ labels by turning each $2\times 2$  block matrix into a vector 
\begin{equation}
B_{\alpha \beta}= \begin{pmatrix}
B_{\alpha \beta}^{(++)}\\
B_{\alpha \beta}^{(+-)}\\
B_{\alpha \beta}^{(-+)}\\
B_{\alpha \beta}^{(--)}
\end{pmatrix},
\end{equation}
and similarly for the gauge field. Since only $B_{\pm \mp}^{(a\,b)}$ mix with the gauge field the diagonalization is identical to that of the fuzzy $S^4$ except that the $B^{(\pm \mp)}_{\alpha \beta}$ fields get a shift in the diagonal coming from $L_{56}^2$. The easiest modes are $\left(B_{\pm \pm}^{(\pm \pm)}\right)_{\boldsymbol{J}}$ which are annihilated by $L_{56}$ and have vanishing matrix elements $T^{\pm \pm}=0$. This means that they do not mix with the gauge field modes and their mass matrix is diagonal:

\begin{equation}
\begin{aligned}
\left(\hat{m}_{\pm \pm}^{(\pm \pm)}\right)^2_{\boldsymbol{J}}&= m^2_{\pm \pm}(L_1, L_2)- \left[C_2^{SO(4)}(J_1, J_2)- C_2^{SO(4)}(L_1, L_2) -C_2^{SO(4)}\left(\frac{1}{2}, \frac{1}{2}\right)\right]\\
\left(\hat{m}_{++}^{(\pm \pm)}\right)^2_{\boldsymbol{J}}&= 4(L_2+2L_1L_2)\\
\left(\hat{m}_{--}^{(\pm \pm)}\right)^2_{\boldsymbol{J}}&= 4(2L_1(L_2+1)+3L_2+2).
\end{aligned}
\end{equation}
The $\left(B_{\pm \pm}^{(\pm \mp)}\right)_{\boldsymbol{J}}$ also don't mix with the gauge field but they have a non-zero eigenvalue for $L_{56}^2$. This amounts in a shift of $-4$ in their masses,
\begin{equation}
\begin{aligned}
\left(\hat{m}_{++}^{(\pm \mp)}\right)^2_{\boldsymbol{J}}&= 4(L_2+2L_1L_2-1)\\
\left(\hat{m}_{--}^{(\pm \mp)}\right)^2_{\boldsymbol{J}}&= 4(2L_1(L_2+1)+3L_2+1).
\end{aligned}
\end{equation}
The remaining set of modes mix with the gauge field, but not between different values of $(a,b)$. This means that we get a set of four $3\times3$ mass matrices; 
\begin{equation}
\begin{aligned}
M^{(ab)}&=\begin{pmatrix} d_{+-}^{(a,b)}&0& -\sqrt{2}\,4 T^{+-}\\
0& d_{-+}^{(a,b)}& -\sqrt{2} \,4T^{-+}\\
-\sqrt{2} \,4T^{+-}& -\sqrt{2}\, 4T^{-+}&  f^{(ab)}
\end{pmatrix},
  \end{aligned}
\end{equation}
with the diagonal entries of the matrices being
\begin{equation}
\begin{aligned}
 d_{+-}^{(\pm,\pm)}&= 4(L_2(2L_1+3)+1)\\
  d_{-+}^{(\pm,\pm)}&= 4(2 L_1+1)(L_2+1)\\
  d_{\pm\mp}^{(\pm,\mp)}&=  d_{\pm\mp}^{(\pm,\pm)}-4\\
  f^{\pm \pm}&= 4(2L_2(L_1+1)+L_1)\\
   f^{\pm \mp}&= f^{\pm \pm}-4.
  \end{aligned}
\end{equation}
The analysis for the off-diagonal blocks is identical up to replacements of the representations $(L_1,L_2)$ to $\mathcal{R}$.
\section{Dimensional Reduction of Spinors}
Now we review some of the details related to the computation of the fermion propagators in various defect backgrounds. For higher codimensions the 4d Dirac operator splits into an $AdS$ factor and a sphere factor. In order to perform this splitting we need to reduce the 4d spinor indices in to the appropriate spinors on $AdS$ and then perform a KK reduction on the sphere. `For $\mathcal{N}=4$ SYM it is useful to start with a Majorana-Weyl in ten dimensions and dimensionally reduce to four dimensions. The 32 component Majorana-Weyl spinor can be decomposed into four 4d Dirac spinors \cite{Ishiki:2006rt}:
\begin{equation}
\Psi= \begin{pmatrix}
\psi^A_+\\
\psi_{-A}
\end{pmatrix}, \;\; A=1,2,3,4,
\end{equation}
where $\psi_{- A}$ is the 4d charge conjugate of $\psi^A_+$. The 10d gamma matrices can be decomposed as
\begin{equation}
\begin{aligned}
\Gamma^\mu= \gamma^\mu \otimes 1_8, \;\;\; \Gamma^{AB}= \gamma_5
\otimes \begin{pmatrix}
0&-\tilde{\rho}^{AB}\\
\rho^{AB}&0
\end{pmatrix}
\end{aligned}
\end{equation}
where $\gamma^\mu$ are 4d gamma matrices, and the matrices $\rho, \tilde{\rho}$ can be chosen to be of the form
\begin{equation}
(\rho^{AB})_{CD}= \delta^A_C \delta^B_D- \delta^A_D \delta^B_C,\;\;\; (\tilde{\rho}^{AB})^{CD}= \epsilon^{ABCD}.
\end{equation}
The reduction to 3d can be done by choosing a basis of gamma matrices such as 
\begin{equation}
\begin{aligned}
    \gamma^0&=  \sigma_1\otimes (i \sigma_1)\\
\gamma^1&=  \sigma_1\otimes \sigma_2\\
\gamma^2&= \sigma_1\otimes \sigma_3\\
\gamma^3&=  \sigma_3\otimes \sigma_2\\
\mathcal{C}&=\gamma^1.
\end{aligned}
\end{equation}
Each 4d Dirac spinor decomposes into a pair of 3d Dirac fermions 

\begin{equation}
\psi^A= \begin{pmatrix}
\lambda^A\\
\chi^A
\end{pmatrix}.
\end{equation}
The kinetic terms for each 4d Dirac fermion reduces to 
\begin{equation}
    S= \int d\psi\int d^3x \sqrt{g_3} \,\frac{i}{2} \left( \
   \left(\bar{\chi} \slashed{D}_{3} \chi + \bar{\lambda} \slashed{D}_3 \lambda \right) + \bar{\chi} \partial_\psi \lambda- \bar{\lambda} \partial_\psi \chi\right)
\end{equation}
where $g_3$ and $\slashed{D}_3$ are the Dirac operator and metric of Poincare $AdS_3$ respectively.

 The reduction from 4d to 2d may be done using the following basis
\begin{equation}
\begin{aligned}
\gamma^{a}&= \sigma^{a+1}\otimes 1_2, \;\;\; a=0,1\\
\gamma^i&= \sigma^3\otimes \sigma^{i},\;\;\; i=2,3.
\end{aligned}
\end{equation}
The spinors decompose as tensors 

\begin{equation}
\psi^A= \lambda^A\otimes \xi,
\end{equation}
where $\xi$ are eigenspinors of the Dirac operator on $S^2$.



\bibliographystyle{JHEP}
	\cleardoublepage
	
\renewcommand*{\bibname}{References}

\bibliography{CodimD}
\end{document}